# Oracularization and Two-Prover One-Round Interactive Proofs against Nonlocal Strategies

Tsuyoshi Ito[a*] Hirotada Kobayashi[b c †] Keiji Matsumoto[b c †]

[a]School of Computer Science
McGill University, Montreal, Quebec, Canada

[b]Principles of Informatics Research Division
National Institute of Informatics, Tokyo, Japan

[c]Quantum Computation and Information Project
Solution Oriented Research for Science and Technology
Japan Science and Technology Agency, Tokyo, Japan

2 October 2008

**Abstract**

A central problem in quantum computational complexity is how to prevent entanglement-assisted cheating in multi-prover interactive proof systems. It is well-known that the standard *oracularization* technique completely fails in some proof systems under the existence of prior entanglement. This paper studies two constructions of *two-prover one-round* interactive proof systems based on oracularization. First, it is proved that the two-prover one-round interactive proof system for PSPACE by Cai, Condon, and Lipton still achieves exponentially small soundness error in the existence of prior entanglement between dishonest provers (and more strongly, even if dishonest provers are allowed to use arbitrary no-signaling strategies). It follows that, unless the polynomial-time hierarchy collapses to the second level, two-prover systems are still advantageous to single-prover systems even when only malicious provers can use quantum information. Second, it is proved that the two-prover one-round interactive proof system obtained by oracularizing a three-query probabilistically checkable proof system becomes sound in a weak sense even against dishonest entangled provers with the help of a dummy question. As a consequence, every language in NEXP has a two-prover one-round interactive proof system of perfect completeness, albeit with exponentially small gap between completeness and soundness, in which each prover responds with only two bits. In other words, it is NP-hard to approximate within an inverse-polynomial the value of a classical two-prover one-round game, even when provers are entangled and each sends a two-bit answer to a verifier.

*Supported by postdoctoral fellowship from the Fonds québécois de la recherche sur la nature et les technologies and the Natural Sciences and Engineering Research Council of Canada.

†Partially supported by the Grant-in-Aid for Scientific Research (B) No. 18300002 of the Ministry of Education, Culture, Sports, Science and Technology of Japan.

# 1 Introduction

## 1.1 Background

Interactive proof systems [3, 16] are a communication model between a polynomial-time probabilistic verifier and a computationally unbounded prover. The prover attempts to convince the verifier that a given input string satisfies some property, while the verifier tries to verify the validity of the assertion of the prover. It is well-known that the model exactly characterizes $\mathrm{PSPACE}$ [24, 27], even with a *public-coin* verifier [17, 28]. Multi-prover interactive proof systems were introduced by Ben-Or, Goldwasser, Kilian, and Wigderson [6] as an important generalization of interactive proof systems, originally for a cryptographic purpose. In this model, a verifier communicates with multiple provers, who are not allowed to communicate with each other. The model turned out to be surprisingly powerful as Babai, Fortnow, and Lund [4] proved that it exactly characterizes $\mathrm{NEXP}$. Together with the older result by Fortnow, Rompel, and Sipser [14], this gave the first step towards the theory of inapproximability and probabilistically checkable proof (PCP) systems [12, 2, 1]. Of particular interests was the power of *two-prover one-round* interactive proof systems. It was already proved in the first paper on multi-prover interactive proofs [6] that two-prover systems are as powerful as general multi-prover systems. Cai, Condon, and Lipton [7] showed that every language in $\mathrm{PSPACE}$ has a two-prover one-round interactive proof system of perfect completeness with exponentially small error in soundness. The combination of the results in Refs. [14, 4] implicitly showed that every language in $\mathrm{NEXP}$ has a two-prover one-round interactive proof system of perfect completeness, but with soundness error only bounded away from one by an inverse-polynomial. Feige [11] improved this to constant soundness error. Finally, Feige and Lovász [13] proved that every language in $\mathrm{NEXP}$ has a two-prover one-round interactive proof system of perfect completeness with exponentially small error in soundness. Thus, the most restricted case of multi-prover interactive proofs is as powerful as the most general case of them. Later Raz [26] showed the *parallel repetition theorem* for a two-prover one-round system, which implies that parallel repetition of a two-prover one-round system reduces the soundness error exponentially fast. It is stressed that *oracularization*, a method that a verifier uses the second prover to force the functional behavior on the first prover, plays essential roles in all of these results except for the parallel repetition theorem.

From a game theoretic viewpoint, multi-prover interactive proof systems can be viewed as cooperative games with imperfect information played by provers and a verifier. More precisely, a $k$-prover $r$-round interactive proof system with a fixed input naturally corresponds to an $r$-round game played by $k$ cooperative players (provers) and a referee (a verifier), where the value of the game is exactly the accepting probability of the underlying proof system with the fixed input. For convenience, $r$-round games with imperfect information played by $k$ cooperative players and a referee are called *$k$-prover $r$-round games* in this paper. In physics, the study of quantum nonlocality has a long history (see Ref. [32] for instance), and in particular, quantum nonlocality is known to significantly affect cooperative games with imperfect information. Even if interactions between a referee and players remain classical, sharing prior entanglement among players increases winning probability in some cooperative games [8]. This means that multi-prover interactive proof systems may become weaker if provers are allowed to share prior entanglement. Indeed, Cleve, Høyer, Toner, and Watrous [8] showed that there are two-prover one-round games where unentangled provers cannot win with certainty but entangled provers can. Among others, the example of the Magic Square game implies that the oracularization paradigm no longer works well in some proof systems in the presence of prior entanglement between the two provers.

Recently, several methods were proposed to limit the power of dishonest entangled provers in multi-prover interactive proof systems. Kempe, Kobayashi, Matsumoto, Toner, and Vidick [20] established two methods to modify a classical multi-prover interactive proof system so that dishonest provers cannot cheat perfectly even with prior entanglement. One is to use quantum messages, and the other is to introduce an additional prover. As a result, they proved that, even in the presence of prior entanglement among dishonest provers, every language in $\mathrm{NEXP}$ has (a) *quantum two-prover* one-round and (b) *classical three-prover* one-round interactive proof systems of perfect completeness, although the proved gap between completeness and soundness is exponentially small.

They asked whether or not a result similar to these holds even in the case of *classical two-prover* one-round. Ito, Kobayashi, Preda, Sun, and Yao [19] proved the existence of a proof system with such properties even with a classical three-prover one-round *binary* interactive proof system, where each prover answers only one bit, and even against *commuting-operator provers*, a model of provers based on the work by Tsirelson [30]. Commuting-operator provers are at least as powerful as usual entangled provers, and capture the case where provers share an entangled state of infinite dimension. Kempe, Kobayashi, Matsumoto, Toner, and Vidick [20] also analyzed the classical two-prover one-round interactive proof system which is obtained by applying oracularization to the public-coin single-prover multi-round interactive proof system, and proved that $\mathrm{PSPACE}$ has classical two-prover one-round interactive proof systems of perfect completeness with soundness error bounded away from one by an inverse-polynomial even against entangled provers. Since the parallel repetition theorem is not known to hold when the provers are entangled, it remained open if $\mathrm{PSPACE}$ has classical two-prover one-round interactive proof systems that achieve arbitrarily small errors even with entangled provers. In fact, the resulting proof system after parallel repetition is exactly the one for which Cai, Condon, and Lipton [7] achieved exponentially small soundness error against unentangled provers. Recently, Holenstein [18] proved that the parallel repetition theorem holds even for two-prover one-round interactive proof systems with no-signaling provers.

## 1.2 Main results

This paper presents two results on the power of (classical) two-prover one-round interactive proof systems against dishonest entangled provers. Let $\mathrm{poly}$ be the set of polynomially-bounded functions that are computable in polynomial time, and let $\mathrm{MIP}^*_{c,s}(2,1)$ and $\mathrm{MIP}^{\mathrm{ns}}_{c,s}(2,1)$ be the classes of languages having two-prover one-round interactive proof systems with completeness at least $c$ and soundness at most $s$, where the provers are allowed to share prior entanglement and to use arbitrary no-signaling strategies, respectively.

First, it is proved that the two-prover one-round interactive proof system for $\mathrm{PSPACE}$ by Cai, Condon and Lipton [7] still achieves exponentially small soundness error against dishonest entangled provers, and more strongly, even against dishonest no-signaling provers. Actually, it is crucial to consider the soundness against no-signaling provers since the proof uses the parallel repetition theorem of two-prover one-round games with no-signaling provers due to Holenstein [18].

**Theorem 1.** *Every language in* $\mathrm{PSPACE}$ *has a two-prover one-round interactive proof system that has perfect completeness with honest unentangled provers and exponentially small soundness error against dishonest no-signaling provers. In particular, for any* $p \in \mathrm{poly}$,

$$\mathrm{PSPACE} \subseteq \mathrm{MIP}^*_{1,2^{-p}}(2,1) \cap \mathrm{MIP}^{\mathrm{ns}}_{1,2^{-p}}(2,1).$$

To prove Theorem 1, we start with a public-coin single-prover interactive proof system for $\mathrm{PSPACE}$, and oracularize it to obtain a two-prover one-round interactive proof system, as done by Kempe, Kobayashi, Matsumoto, Toner, and Vidick [20]. In the constructed system, the verifier sends the whole sequence of $r$ public-coin questions in the original system to the first prover at one time, where $r$ is the number of rounds in the original system. The verifier then uniformly chooses a round $k \in \{1,\dots,r\}$, and sends only the first $k$ public-coin questions to the second prover. The verifier checks that (i) the $r$ public-coin questions and the $r$ answers from the first prover form an accepting conversation in the original system (SIMULATION TEST), and (ii) the $k$ answers from the second prover are all consistent with the first $k$ answers from the first prover (CONSISTENCY TEST). It is proved that this two-prover one-round interactive proof system has soundness error bounded away from one by an inverse-polynomial even against dishonest no-signaling provers (previously, the soundness was shown only against dishonest entangled provers in Ref. [20]). Now this proof system is repeated in parallel polynomially many times. The resulting proof system is exactly the one analyzed by Cai, Condon, and Lipton [7] in the classical setting with unentangled provers. The parallel repetition theorem with no-signaling provers [18] shows that the resulting system achieves exponentially small soundness error. Since entangled provers and commuting-operator provers are special cases of no-signaling provers, the same soundness holds even against provers of these types.

When proving soundness, we construct a strategy of the single-prover system from a no-signaling strategy of the oracularized system. This can be viewed as a *rounding* in the terminology in Ref. [20]. The original rounding in Ref. [20] uses post-measurement quantum states. Since there is no notion corresponding to post-measurement states in the case of no-signaling provers, this paper presents a new method of rounding, which is essential for proving a bound for no-signaling strategies.

Theorem 1 implies that unless $\mathrm{AM} = \mathrm{PSPACE}$, in particular unless the polynomial-time hierarchy collapses to the second level, two-prover systems are still advantageous to single-prover systems even when only dishonest provers can use quantum information. To the best knowledge of the authors, this is the first nontrivial lower bound on the power of two-prover one-round interactive proof systems with entangled and no-signaling provers.

Since any two-prover one-round interactive proof system with no-signaling provers is efficiently simulatable by solving a linear program of exponential size, Theorem 1 implies that $\mathrm{PSPACE} \subseteq \mathrm{MIP}^{\mathrm{ns}}_{c,s}(2,1) \subseteq \mathrm{EXP}$ for any polynomial-time computable functions $c, s$ satisfying $c > s$.

Next, it is proved that every language in NEXP has a (classical) two-prover one-round interactive proof system of perfect completeness with soundness bounded away from one even against entangled provers (actually against commuting-operator provers), although the gap between completeness and soundness is exponentially small. This affirmatively answers the question posed in Ref. [20]. More precisely, the following is proved.

**Theorem 2.** *Every language in* $\mathrm{NEXP}$ *has a two-prover one-round interactive proof system satisfying the following properties:*

*(i) Each prover responds with two bits.*

*(ii) It has perfect completeness with honest unentangled provers.*

*(iii) It has soundness error at most* $1 - 2^{-p}$ *against dishonest commuting-operator provers (hence also against dishonest entangled provers) for some* $p \in \mathrm{poly}$ *depending on the language.*

*(iv) It has constant soundness against dishonest unentangled provers.*

*In particular, properties (ii) and (iii) imply*

$$\mathrm{NEXP} \subseteq \bigcup_{p \in \mathrm{poly}} \mathrm{MIP}^{*}_{1,1-2^{-p}}(2,1).$$

Note that this is in contrast to the case with no-signaling provers, where for any $p \in \mathrm{poly}$, $\mathrm{MIP}^{\mathrm{ns}}_{1,1-2^{-p}}(2,1) \subseteq \mathrm{EXP}$.

To prove Theorem 2, we start with a nonadaptive three-query PCP system of perfect completeness, and oracularize it to obtain a two-prover one-round interactive proof system. In the usual oracularization, the verifier sends all the three questions in the original PCP system to the first prover to simulate the original PCP system (SIMULATION TEST), and one of the three chosen uniformly at random to the second prover to check the consistency with the first prover (CONSISTENCY TEST). However, this is not sufficient for our purpose, since the example of the Magic Square game implies that the system is perfectly cheatable by entangled provers when the underlying PCP system has a very bad structure. The main idea to overcome this difficulty is that the verifier sends a *dummy* question to the second prover in addition to the original question used to check the consistency — the verifier randomly chooses one dummy question from all the possible questions in the original PCP system, and sends both of the original question for the CONSISTENCY TEST and this dummy question to the second prover (of course, without revealing which one is dummy). The verifier checks the consistency using the answer for the real question, and just ignores the answer for the dummy question. The intuition behind this is as follows. In order to cheat the proof system, the answers from the first prover must be highly nonlocally correlated to the answer to the real question from the second prover. However, the second prover would be forced to use entanglement even for the dummy question, since he does not know which one is dummy. Now the answers from the first prover could not be highly nonlocally correlated only to the answer to the real question from the second prover, due to *monogamy of quantum correlations* [29]. It turns out that this dummy question is sufficiently helpful to at least prove the following two properties whenever the provers use a commuting-operator strategy that passes the CONSISTENCY TEST with very high probability: (a) for each measurement operator $M$ used by one prover, there is a measurement operator by the other prover whose effect is close to $M$ when applied to the shared state, and (b) the measurement operators by each prover are pairwise almost commuting when applied to the shared state. These two properties are also key in the proof in Ref. [20], and our new contribution lies in presenting a method to derive these properties by using only two provers and classical messages. Once these properties are obtained, one can use a rounding technique similar to Ref. [20] to construct a (randomized) PCP proof of the original system from the entangled strategy with these two properties so that the accepting probability with such entangled provers is not very far from that in the original PCP system when the constructed PCP proof is given. This implies that the first prover fails in the SIMULATION TEST with high probability whenever the provers passes the CONSISTENCY TEST with very high probability, and proves the soundness bound. Since $\mathrm{NEXP}$ has nonadaptive three-query PCP systems of perfect completeness with constant soundness error [5] (using polynomially many random bits), the above argument essentially shows Theorem 2. Finally, the property (i) of Theorem 2 that two-bit-answer systems are sufficient follows from the nonadaptive three-query PCP system naturally induced by the scaled-up version of the inapproximability result of 1-IN-3 3SAT by Khanna, Sudan, Trevisan, and Williamson [22] — it is $\mathrm{NEXP}$-hard to distinguish whether a given instance of SUCCINCT 1-IN-3 3SAT is satisfiable or at most a constant fraction of its clauses is simultaneously satisfiable. Since the underlying PCP system is for SUCCINCT 1-IN-3 3SAT, there are at most three possibilities for the answer of the first prover, which can be encoded in one trit (and thus in two bits).

The same argument can be applied to a nonadaptive three-query PCP system for $\mathrm{NP}$, which proves the $\mathrm{NP}$-hardness of approximating the entangled value of a two-prover one-round game. For a two-prover one-round game $G$, let $w_{\mathrm{unent}}(G)$, $w_{\mathrm{ent}}(G)$, and $w_{\mathrm{com}}(G)$ denote the unentangled, entangled, and commuting-operator values of $G$, respectively, i.e., the values of the game $G$ when played optimally by unentangled, entangled, and commuting-operator provers.

**Corollary 3.** *The following promise problem is* $\mathrm{NP}$*-complete.*

*Input: A two-prover one-round game $G$ with two-bit answers and an integer $k \geq 1$, where $G$ is represented as the complete tables of the probability distribution $\pi$ over two questions $q_1, q_2$ and of the predicate $R$ of the referee, and $k$ is represented in unary.*

*Yes-instance:* $w_{\mathrm{unent}}(G) = 1$.

*No-instance:* $w_{\mathrm{com}}(G) \leq 1 - 1/k$.
*(This condition is at least as strong as the condition $w_{\mathrm{ent}}(G) \leq 1 - 1/k$.)*

Corollary 3 implies that, unless $\mathrm{P} = \mathrm{NP}$, there does not exist a fully polynomial time approximation scheme (FPTAS) to approximate $w_{\mathrm{ent}}(G)$ or $w_{\mathrm{com}}(G)$ where $G$ is a classical two-prover one-round game with two-bit answers. In particular, the entangled value of such games cannot be represented by a semidefinite program of polynomial size, unlike the so-called XOR games [30, 31, 8]. Corollary 3 is also in contrast to the case of two-prover one-round games with *binary* answers, where it is efficiently decidable whether the entangled value of the game is one or not [8].

*Remark.* In fact, the answer from one of the provers can be limited to one trit in Theorem 2 and Corollary 3. Also, as will be presented in Section 4, the games used in the proof of Theorem 2 and Corollary 3 can be restricted to two-to-two games. For a two-to-two game $G$, there is a polynomial-time algorithm based on semidefinite programming that decides whether $w_{\mathrm{ent}}(G) = 1$ or $w_{\mathrm{ent}}(G) < 1/20$ [21].

## 2 Preliminaries

### 2.1 Notations and useful facts

We assume familiarity with the quantum formalism, including the definitions of pure and mixed quantum states, admissible transformations (completely-positive trace-preserving mappings), measurements, and positive operator-valued measures (POVMs) (all of which are discussed in detail in Refs. [25, 23], for instance). This section summarizes some of the notions and notations that are used in this paper.

Throughout this paper, let $\mathbb{N}$ and $\mathbb{Z}^+$ denote the sets of positive and nonnegative integers, respectively. For every $N \in \mathbb{N}$, let $[N]$ denote the set of all positive integers at most $N$. Let $\mathrm{poly}$ be the set of functions $p\colon \mathbb{Z}^+ \to \mathbb{N}$ such that $1^{p(n)}$ is computable in time polynomial in $n$. For any finite set $S$, we associate with $S$ the Hilbert space $\mathbb{C}^S$ of dimension $|S|$ whose orthonormal basis is given by $\{|s\rangle\colon s \in S\}$. For any Hermitian operator $A$, the *trace norm* of $A$ is defined as $\|A\|_{\mathrm{tr}} = \mathrm{tr}\sqrt{A^\dagger A}$. For any Hilbert space $\mathcal{H}$, $\mathbf{D}(\mathcal{H})$ denotes the set of density operators over $\mathcal{H}$. For any Hilbert spaces $\mathcal{H}$ and $\mathcal{K}$, $\mathbf{T}(\mathcal{H}, \mathcal{K})$ denotes the set of admissible transformations from $\mathbf{D}(\mathcal{H})$ to $\mathbf{D}(\mathcal{K})$.

For any Hilbert space $\mathcal{H}$ and any finite sets $S$ and $T$, two POVMs $\boldsymbol{M} = \{M_s\}_{s\in S}$ and $\boldsymbol{N} = \{N_t\}_{t\in T}$ over $\mathcal{H}$ *mutually commute* if $[M_s, N_t] = M_s N_t - N_t M_s = 0$ for any $s \in S$ and $t \in T$.

For any finite set $S$ and for any probability distributions $p$ and $q$ over $S$, the *statistical difference* between $p$ and $q$ is defined as $\mathrm{SD}(p, q) = \frac{1}{2}\sum_{s\in S} |p(s) - q(s)|$. For any Hilbert space $\mathcal{H}$ and for any density operators $\rho$ and $\sigma$ in $\mathbf{D}(\mathcal{H})$, the *trace distance* between $\rho$ and $\sigma$ is defined as $D(\rho, \sigma) = \frac{1}{2}\|\rho - \sigma\|_{\mathrm{tr}}$. The following properties will be used in the subsequent sections.

**Fact 1.** For any Hilbert space $\mathcal{H}$ and for any pure quantum states $|\phi\rangle$ and $|\psi\rangle$ in $\mathcal{H}$,

$$D(|\phi\rangle\langle\phi|, |\psi\rangle\langle\psi|) = \sqrt{1 - |\langle\phi|\psi\rangle|^2}.$$

**Fact 2.** For any Hilbert spaces $\mathcal{H}$ and $\mathcal{K}$, any density operators $\rho$ and $\sigma$ in $\mathbf{D}(\mathcal{H})$, and any admissible transformation $T$ in $\mathbf{T}(\mathcal{H}, \mathcal{K})$,

$$D(T(\rho), T(\sigma)) \leq D(\rho, \sigma).$$

In particular, Fact 2 implies that, for any Hilbert space $\mathcal{H}$, any finite set $S$, and any POVM $\boldsymbol{M} = \{M_s\}_{s\in S}$ over $\mathcal{H}$, $D(\rho, \sigma) \geq \frac{1}{2}\sum_{s\in S} |\mathrm{tr}M_s\rho - \mathrm{tr}M_s\sigma| = \mathrm{SD}(p^{\boldsymbol{M}}(\rho), p^{\boldsymbol{M}}(\sigma))$, where $p^{\boldsymbol{M}}(\rho)$ and $p^{\boldsymbol{M}}(\sigma)$ are the probability distributions over $S$ naturally induced by performing the measurement associated with $\boldsymbol{M}$ on $\rho$ and $\sigma$, respectively.

In what follows, $D(|\phi\rangle\langle\phi|, |\psi\rangle\langle\psi|)$ for pure states $|\phi\rangle$ and $|\psi\rangle$ in $\mathcal{H}$ is often simply denoted by $D(|\phi\rangle, |\psi\rangle)$, and $(D(\rho, \sigma))^2$ is denoted by $D^2(\rho, \sigma)$.

### 2.2 Games

This subsection introduces three types of games that are discussed in this paper: two-prover one-round games, nonadaptive-query single-prover multi-round games, and nonadaptive three-query PCP games.

The notion of games are intimately related to the notion of interactive proof and probabilistically checkable proof systems. We assume familiarity with these proof systems, see, e.g., Refs. [10, 15].

For convenience, a family $\{\theta_s\}_{s\in S}$ of probability distributions over a set $T$ is often simply denoted by $\theta$, sometimes even without mentioning the underlying sets $S$ and $T$ when it is not confusing. Also, the probability $\theta_s(t)$ is often denoted by $\theta(t \mid s)$.

#### Two-prover one-round games

*Two-prover one-round games* are cooperative simultaneous games played by two cooperative players $P_1$ and $P_2$ with a referee $V$. In what follows, the referee and players are called the verifier and provers, respectively, for the terminological consistency with multi-prover interactive proofs. A two-prover one-round game $G$ is specified by a six-tuple $(Q_1, Q_2, A_1, A_2, R, \pi)$, where $Q_1$, $Q_2$, $A_1$, and $A_2$ are finite sets, $R\colon Q_1 \times Q_2 \times A_1 \times A_2 \to \{0,1\}$ is a predicate, and $\pi$ is a probability distribution over $Q_1 \times Q_2$. The verifier $V$ chooses a pair $(q_1, q_2)$ of questions from $Q_1 \times Q_2$ according to $\pi$, and sends $q_i$ to the prover $P_i$ for each $i \in \{1,2\}$. Each $P_i$ answers $a_i \in A_i$, and the provers win the game if and only if $R(q_1, q_2, a_1, a_2) = 1$. Following convention, $R(q_1, q_2, a_1, a_2)$ is denoted by $R(a_1, a_2 \mid q_1, q_2)$ in this paper.

A *strategy* of the provers in a two-prover one-round game $G = (Q_1, Q_2, A_1, A_2, R, \pi)$ is a family $\theta = \{\theta_{q_1,q_2}\}_{(q_1,q_2)\in Q_1\times Q_2}$ of probability distributions over $A_1 \times A_2$ indexed by each element in $Q_1 \times Q_2$. This can be interpreted as the provers jointly answering $(a_1, a_2) \in A_1 \times A_2$ with probability $\theta(a_1, a_2 \mid q_1, q_2) = \theta_{q_1,q_2}(a_1, a_2)$ when the verifier asks $q_1 \in Q_1$ to the first prover and $q_2 \in Q_2$ to the second prover. Depending on the resources the provers can use, this paper considers the following four classes of strategies of the provers.

The smallest one is the class of *unentangled* (or *classical*) strategies in which the provers are allowed to share any classical random source before the game starts. Each prover can use his private random source also, and can be an arbitrary function depending on the question he receives and the shared and private random sources.

The second smallest class is that of *entangled* strategies in which the provers are allowed to share an arbitrary quantum state in a Hilbert space $\mathcal{P}_1 \otimes \mathcal{P}_2$, where each $\mathcal{P}_1$ and $\mathcal{P}_2$ can be arbitrarily large Hilbert space of finite dimension. Each prover $P_i$ can perform an arbitrary POVM measurement over $\mathcal{P}_i$ that depends on the question he receives, and he answers with this measurement outcome.

A more general one is the class of *commuting-operator* strategies in which the provers are again allowed to share any quantum state in a Hilbert space $\mathcal{P}$, but now $\mathcal{P}$ can be infinite-dimensional. Each prover $P_i$ can perform an arbitrary POVM measurement over $\mathcal{P}$ as long as the measurement by $P_1$ commutes with that by $P_2$.

Finally, the largest one of the four is the class of *no-signaling* strategies in which the provers are allowed to do anything as long as their behavior viewed from the outside does not imply signaling between the two provers. Formally, a strategy $\theta$ is no-signaling if for any $i \in \{1,2\}$ and for any question $q_i \in Q_i$, the marginal distribution

$$\sum_{a_{3-i}\in A_{3-i}} \theta(a_1, a_2 \mid q_1, q_2)$$

does not depend on the choice of $q_{3-i}$.

The *winning probability* or the *accepting probability* of a strategy $\theta$ is given by

$$\sum_{(q_1,q_2)\in Q_1\times Q_2} \pi(q_1, q_2) \sum_{(a_1,a_2)\in A_1\times A_2} \theta(a_1, a_2 \mid q_1, q_2) R(a_1, a_2 \mid q_1, q_2).$$

By convexity argument, the classical random source is not necessary when considering the optimal unentangled strategy. Similarly, it is assumed without loss of generality that the shared state is pure and the measurement by the provers are projective in entangled and commuting-operator strategies.

Corresponding to the four classes of strategies, a game $G$ has four values: the *unentangled value* $w_{\mathrm{unent}}(G)$, the *entangled value* $w_{\mathrm{ent}}(G)$, the *commuting-operator value* (or the *field-theoretic value* in the terminology used in Ref. [9]) $w_{\mathrm{com}}(G)$, and the *no-signaling value* $w_{\mathrm{ns}}(G)$. All of these are defined as the supremum of the winning probability in $G$ over all strategies in that class. The unentangled and the no-signaling values are known to be attainable. Clearly,

$$0 \le w_{\mathrm{unent}}(G) \le w_{\mathrm{ent}}(G) \le w_{\mathrm{com}}(G) \le w_{\mathrm{ns}}(G) \le 1.$$

Holenstein [18] showed the following parallel repetition theorem for the no-signaling values of two-prover one-round games.

**Theorem 4** ([18]). *There exist positive constants $c$ and $\alpha$ such that for any two-prover one-round game $G$ and any $n \in \mathbb{N}$, the two-prover one-round game $G^n$ obtained by the $n$-fold parallel repetition of $G$ satisfies $w_{\mathrm{ns}}(G^n) \le (1 - c(1 - w_{\mathrm{ns}}(G))^{\alpha})^n$.*

### Nonadaptive-query single-prover multi-round games

*Nonadaptive-query single-prover multi-round games* are games played by a single player (a prover) $P$ with a referee (a verifier) $V$. A nonadaptive-query single-prover $r$-round game $G$ is specified by a quadruple $(Q, A, R, \pi)$, where $Q$ and $A$ are finite sets, $R\colon Q^r \times A^r \to \{0,1\}$ is a predicate, and $\pi$ is a probability distribution over $Q^r$. At the beginning of the game $G$, the verifier $V$ chooses an $r$-tuple $(q_1, \dots, q_r)$ of questions from $Q^r$ according to $\pi$. At the $j$th round for each $1 \leq j \leq r$, $V$ sends $q_j$ to the prover $P$, and $P$ answers $a_j \in A$. The prover wins the game if and only if $R(q_1, \dots, q_r, a_1, \dots, a_r) = 1$. Similarly to the case of two-prover one-round games, $R(q_1, \dots, q_r, a_1, \dots, a_r)$ is denoted by $R(a_1, \dots, a_r \mid q_1, \dots, q_r)$ in this paper.

A *strategy* of the prover in a nonadaptive-query single-prover $r$-round game $G = (Q, A, R, \pi)$ is a sequence $\boldsymbol{\theta} = (\theta^{(1)}, \dots, \theta^{(r)})$ of $r$ families of probability distributions over $A$, where $\theta^{(j)}$ is a family indexed by each element in $Q^j \times A^{j-1}$. This can be interpreted as the prover answering $a_j \in A$ at the $j$th round, for each $1 \leq j \leq r$, with probability $\theta^{(j)}(a_j \mid q_1, \dots, q_j, a_1, \dots, a_{j-1})$ when the conversation between the verifier and the prover so far forms a sequence $(q_1, a_1, \dots, q_{j-1}, a_{j-1}, q_j) \in (Q \times A)^{j-1} \times Q$.

A strategy $\boldsymbol{\theta}$ naturally induces a family $\{\theta_{q_1,\dots,q_r}\}_{(q_1,\dots,q_r)\in Q^r}$ of probability distributions over $A^r$ defined by

$$\theta_{q_1,\dots,q_r}(a_1, \dots, a_r) = \prod_{j=1}^{r} \theta^{(j)}(a_j \mid q_1, \dots, q_j, a_1, \dots, a_{j-1}),$$

and as before, $\theta_{q_1,\dots,q_r}(a_1, \dots, a_r)$ is denoted by $\theta(a_1, \dots, a_r \mid q_1, \dots, q_r)$. The *winning probability* or the *accepting probability* of the strategy is given by

$$\sum_{q_1,\dots,q_r \in Q} \pi(q_1, \dots, q_r) \sum_{a_1,\dots,a_r \in A} \theta(a_1, \dots, a_r \mid q_1, \dots, q_r) R(a_1, \dots, a_r \mid q_1, \dots, q_r). \tag{1}$$

The *value* $w(G)$ of the game $G$ is the maximum winning probability over all strategies of the prover.

It is noted that we allow the strategies of the prover to be probabilistic so that analyses in Section 3 become easier, although there always exists a deterministic optimal strategy.

### Nonadaptive three-query PCP games

*Nonadaptive three-query PCP games* are one-way communication games played by a single player (a prover) $P$ with a referee (a verifier) $V$. A nonadaptive three-query PCP game $G$ is specified by a quadruple $(Q, A, R, \pi)$, where $Q$ is a positive integer, $A$ is a finite set,[1] $R\colon [Q]^3 \times A^3 \to \{0,1\}$ is a predicate, and $\pi$ is a probability distribution over the set $\{(q_1, q_2, q_3)\colon 1 \leq q_1 < q_2 < q_3 \leq Q\} \subseteq [Q]^3$. The game starts with the prover $P$ sending a string $\Pi \in A^Q$ on alphabet $A$ of length $Q$. The verifier $V$ chooses three positions $q_1$, $q_2$, and $q_3$, $1 \leq q_1 < q_2 < q_3 \leq Q$, according to $\pi$, and reads the three letters $a_1$, $a_2$, and $a_3$ of $\Pi$ in the corresponding positions. The prover wins the game if and only if $R(q_1, q_2, q_3, a_1, a_2, a_3) = 1$. As before, $R(q_1, q_2, q_3, a_1, a_2, a_3)$ is denoted by $R(a_1, a_2, a_3 \mid q_1, q_2, q_3)$.

A *strategy* of the prover in a nonadaptive three-query PCP game $G = (Q, A, R, \pi)$ is a probability distribution $\theta$ over $A^Q$. This can be interpreted as the prover preparing a string $\Pi \in A^Q$ with probability $\theta(\Pi)$. For notational convenience, the $q$th letter of $\Pi$ is denoted by $\Pi[q]$, for each $1 \leq q \leq Q$.

The strategy $\theta$ naturally induces a family of probability distributions over $A^3$ indexed by $[Q]^3$ defined as

$$\theta(a_1, a_2, a_3 \mid q_1, q_2, q_3) = \sum_{\substack{\Pi \in A^Q \\ \Pi[q_i] = a_i \ (i=1,2,3)}} \theta(\Pi).$$

[1] Some might prefer to fix $A$ to the binary answer $\{0,1\}$ to make the definition consistent with the name, since "nonadaptive three-query PCP systems" usually refer to such PCP systems with the binary alphabet. Actually, in the proofs of Theorem 2 and Corollary 3, we only use the case where $A = \{0,1\}$. However, we use this definition to make it consistent with the definitions of games of other kinds.

The *winning probability* or the *accepting probability* of a strategy $\theta$ is given by

$$\sum_{1\le q_1<q_2<q_3\le Q} \pi(q_1,q_2,q_3) \sum_{\Pi\in A^Q} \theta(\Pi) R\bigl(\Pi[q_1],\Pi[q_2],\Pi[q_3] \mid q_1,q_2,q_3\bigr)$$
$$= \sum_{1\le q_1<q_2<q_3\le Q} \pi(q_1,q_2,q_3) \sum_{a_1,a_2,a_3\in A} \theta(a_1,a_2,a_3 \mid q_1,q_2,q_3) R(a_1,a_2,a_3 \mid q_1,q_2,q_3).$$

The *value* $w(G)$ of the game $G$ is the maximum winning probability over all strategies of the prover.

As in the case of nonadaptive-query single-prover multi-round games, we allow the strategies of the prover to be probabilistic so that analyses in Section 4 become easier, although there always exists a deterministic optimal strategy.

*Remark.* In the definition of nonadaptive three-query PCP games, the order of locations $1,\dots,Q$ is used only to define the unique order of a triple $(q_1,q_2,q_3)$. Relabeling locations does not affect the essential meaning of the game or the value of the game.

## 2.3 Oracularizations

*Oracularization* of a game is a way of transforming the game to a two-prover one-round game, where the second prover is used to force some functional behavior on the first prover.

### Oracularization of nonadaptive-query single-prover multi-round games

For a nonadaptive-query single-prover $r$-round game $G=(Q,A,R,\pi)$, oracularization of $G$ gives the two-prover one-round game $G'=(Q^r,Q_2,A^r,A_2,R',\pi')$, where the finite sets $Q_2$ and $A_2$, the predicate $R'$, and the probability distribution $\pi'$ are specified below.

Let $Q_2=\bigcup_{k\in[r]} Q^k$ and $A_2=\bigcup_{k\in[r]} A^k$. The verifier $V'$ in $G'$ chooses $(q_1,\dots,q_r)\in Q^r$ according to $\pi$ and $k\in[r]$ uniformly at random, and sends $(q_1,\dots,q_r)$ to the first prover and $(q_1,\dots,q_k)$ to the second prover. This specifies the probability distribution $\pi'$ over the set $Q^r\times Q_2$. Upon receiving answers $(a_1,\dots,a_r)\in A^r$ from the first prover and $(a'_1,\dots,a'_{k'})\in A^{k'}$ from the second prover, $V'$ performs the following two tests:

(SIMULATION TEST) $V'$ verifies that $R(a_1,\dots,a_r \mid q_1,\dots,q_r)=1$,

(CONSISTENCY TEST) $V'$ verifies that $k=k'$ and $a_j=a'_j$ for every $j\in[k]$.

The provers win if and only if they pass both of the two test. This specifies the predicate $R'\colon Q^r\times Q_2\times A^r\times A_2\to\{0,1\}$.

### Oracularization of nonadaptive three-query PCP games

For a nonadaptive three-query PCP game $G=(Q,A,R,\pi)$, oracularization of $G$ gives the two-prover one-round game $G'=([Q]^3,[Q],A^3,A,R',\pi')$, where the predicate $R'$ and the probability distribution $\pi'$ are specified below.

The verifier $V'$ in $G'$ chooses three positions $q_1$, $q_2$, and $q_3$, $1\le q_1<q_2<q_3\le Q$, according to $\pi$, and then chooses $q\in\{q_1,q_2,q_3\}$ uniformly at random. $V'$ sends $(q_1,q_2,q_3)$ to the first prover and $q$ to the second prover. This specifies the probability distribution $\pi'$ over the set $[Q]^3\times[Q]$. Upon receiving answers $(a_1,a_2,a_3)\in A^3$ from the first prover and $a\in A$ from the second prover, $V'$ performs the following two tests:

(SIMULATION TEST) $V'$ verifies that $R(a_1,a_2,a_3 \mid q_1,q_2,q_3)=1$,

(CONSISTENCY TEST) $V'$ verifies that $a=a_j$ when $q=q_j$.

The provers win if and only if they pass both of the two tests. This specifies the predicate $R'\colon [Q]^3 \times [Q] \times A^3 \times A \to \{0,1\}$.

The following result is implicit in Ref. [14] (combined with Refs. [4, 6]).

**Theorem 5** ([14, 4, 6])**.** *For a nonadaptive three-query PCP game $G$ and its oracularization $G'$,*

$$w(G) \leq w_{\mathrm{unent}}(G') \leq 1 - \frac{1 - w(G)}{3}.$$

# 3 Two-prover one-round system for PSPACE

This section analyses the two-prover one-round game constructed by oracularizing a nonadaptive-query single-prover multi-round game, and proves Theorem 1 that every language in PSPACE has a two-prover one-round interactive proof system of perfect completeness with exponentially small soundness error against no-signaling provers.

## 3.1 Key lemma

Given a nonadaptive-query single-prover multi-round game $G$, let $G'$ be the two-prover one-round game obtained by oracularizing $G$. As stated in the introduction, Theorem 1 is proved by first relating the value of the original game $G$ with the no-signaling value of the oracularized game $G'$, and then repeating $G'$ in parallel. The main part of the proof of Theorem 1 is to show the following lemma, which relates $w_{\mathrm{ns}}(G')$ with $w(G)$.

**Lemma 6.** *Let $G$ be a nonadaptive-query single-prover $r$-round game, and let $G'$ be the two-prover one-round game obtained by oracularizing $G$. Then,*

$$w_{\mathrm{ns}}(G') \leq 1 - \frac{1 - w(G)}{3r}.$$

The proof of Lemma 6 is deferred to the next subsection. Assuming Lemma 6, it is easy to prove Theorem 1, by using the following celebrated characterization of PSPACE.

**Theorem 7** ([24, 27, 17, 28])**.** *Every language in* PSPACE *has a public-coin single-prover interactive proof system of perfect completeness with soundness error $2^{-p}$ for any $p \in \mathrm{poly}$.*

*Proof of Theorem 1.* Let $L$ be a language in PSPACE. Then, from Theorem 7, $L$ has a public-coin single-prover interactive proof system of perfect completeness with soundness error at most $1/2$. Given an input of length $n$, the system uses $r(n)$ rounds for some $r \in \mathrm{poly}$. The oracularization of this proof system gives a two-prover one-round interactive proof system that clearly achieves perfect completeness with unentangled honest provers. Lemma 6 implies that the soundness error in the constructed two-prover one-round system is at most $1 - \frac{1}{6r}$ against no-signaling dishonest provers. Finally, Theorem 4 shows that, for any $p \in \mathrm{poly}$, there exists $t \in \mathrm{poly}$ such that repeating this system $t$ times in parallel reduces soundness error to at most $2^{-p}$ even against no-signaling dishonest provers, which completes the proof. □

## 3.2 Proof of Lemma 6

Fix a nonadaptive-query single-prover $r$-round game $G = (Q, A, R, \pi)$, and let $G' = (Q^r, Q_2, A^r, A_2, R', \pi')$ be the oracularization of $G$.

For any $\boldsymbol{q} = (q_1, \dots, q_r)$ in $Q^r$, $\boldsymbol{a} = (a_1, \dots, a_r)$ in $A^r$, and $k \in [r]$, let $\boldsymbol{q}_{[1,k]}$ and $\boldsymbol{a}_{[1,k]}$ denote $(q_1, \dots, q_k)$ and $(a_1, \dots, a_k)$, the prefixes of $\boldsymbol{q}$ and $\boldsymbol{a}$ of size $k$, respectively.

Consider any no-signaling strategy $\theta'$ in $G'$, and let $\varepsilon$ be the probability that the provers loses in $G'$ when using the strategy $\theta'$. It is assumed without loss of generality that, in the strategy $\theta'$, the second prover always answers an element in $A^k$ when he receives a question in $Q^k$ (otherwise it only decreases the winning probability of the provers). We shall prove that $\varepsilon \geq \frac{1-w(G)}{3r}$.

Let $\varepsilon_{\text{cons}}$ and $\varepsilon_{\text{sim}}$ be the probabilities that the strategy $\theta'$ fails in CONSISTENCY TEST and SIMULATION TEST, respectively. Obviously, $\varepsilon \geq \max\{\varepsilon_{\text{cons}}, \varepsilon_{\text{sim}}\}$.

For each $\boldsymbol{q} \in Q^r$, let $\varepsilon_{\text{cons}}(\boldsymbol{q})$ be the probability that $\theta'$ fails in CONSISTENCY TEST conditioned on the question to the first prover being $\boldsymbol{q}$. Furthermore, for each $\boldsymbol{q} \in Q^r$ and $k \in [r]$, let $\varepsilon_{\text{cons}}(\boldsymbol{q}, k)$ be the probability that $\theta'$ fails in CONSISTENCY TEST conditioned on the questions to the first and second prover being $\boldsymbol{q}$ and $\boldsymbol{q}_{[1,k]}$. The following relations are obvious:

$$\begin{aligned} \varepsilon_{\text{cons}} &= \sum_{\boldsymbol{q} \in Q^r} \pi(\boldsymbol{q}) \varepsilon_{\text{cons}}(\boldsymbol{q}), \\ \varepsilon_{\text{cons}}(\boldsymbol{q}) &= \frac{1}{r} \sum_{k \in [r]} \varepsilon_{\text{cons}}(\boldsymbol{q}, k), \\ \varepsilon_{\text{cons}}(\boldsymbol{q}, k) &= \sum_{\boldsymbol{a} \in A^r, \boldsymbol{a}' \in A^k \colon \boldsymbol{a}_{[1,k]} \neq \boldsymbol{a}'} \theta'(\boldsymbol{a}, \boldsymbol{a}' \mid \boldsymbol{q}, \boldsymbol{q}_{[1,k]}). \end{aligned}$$

Next, for each $k \in [r]$, let $\varepsilon(k)$ be the probability that $\theta'$ fails in either CONSISTENCY TEST or SIMULATION TEST conditioned on the question to the second prover being in $Q^k$. Again, the following relation clearly holds:

$$\varepsilon = \frac{1}{r} \sum_{k \in [r]} \varepsilon(k) \geq \frac{\varepsilon(r)}{r}. \tag{2}$$

From the no-signaling conditions, for each $\boldsymbol{q} \in Q^r$ and $k \in [r]$, the probability distribution $\alpha_{\boldsymbol{q}}$ over $A^r$ of answers from the first prover when asked $\boldsymbol{q}$ does not depend on $k$ the verifier has chosen. Or equivalently, $\alpha_{\boldsymbol{q}}(\boldsymbol{a}) = \sum_{\boldsymbol{a}' \in A^k} \theta'(\boldsymbol{a}, \boldsymbol{a}' \mid \boldsymbol{q}, \boldsymbol{q}_{[1,k]})$ does not depend on the choice of $k$. Similarly, for each $\boldsymbol{q} = (q_1, \dots, q_r) \in Q^r$ and $k \in [r]$, the probability distribution $\beta_{\boldsymbol{q}_{[1,k]}}$ over $A^k$ of answers from the second prover when asked $\boldsymbol{q}_{[1,k]}$ does not depend on $q_{k+1}, \dots, q_r$. That is, for each $k \in [r]$, each $\boldsymbol{q}_{[1,k]} \in Q^k$, and each $\boldsymbol{a}' \in A^k$, the probability $\beta_{\boldsymbol{q}_{[1,k]}}(\boldsymbol{a}') = \sum_{\boldsymbol{a} \in A^r} \theta'(\boldsymbol{a}, \boldsymbol{a}' \mid \boldsymbol{q}, \boldsymbol{q}_{[1,k]})$ does not depend on $q_{k+1}, \dots, q_r$. For each $\boldsymbol{q} \in Q^r$ and $k \in [r]$, let $\alpha_{\boldsymbol{q},k}$ be the probability distribution over $A^k$ defined by $\alpha_{\boldsymbol{q},k}(\boldsymbol{a}') = \sum_{\boldsymbol{a} \in A^r \colon \boldsymbol{a}_{[1,k]} = \boldsymbol{a}'} \alpha_{\boldsymbol{q}}(\boldsymbol{a})$. It is proved that $\alpha_{\boldsymbol{q},k}$ and $\beta_{\boldsymbol{q}_{[1,k]}}$ are close to each other.

**Claim 1.** *For each $\boldsymbol{q} \in Q^r$ and $k \in [r]$,*

$$\mathrm{SD}(\alpha_{\boldsymbol{q},k}, \beta_{\boldsymbol{q}_{[1,k]}}) \leq \varepsilon_{\text{cons}}(\boldsymbol{q}, k).$$

*Proof.* By definitions,

$$\alpha_{\boldsymbol{q},k}(\boldsymbol{a}') = \sum_{\boldsymbol{a} \in A^r \colon \boldsymbol{a}_{[1,k]} = \boldsymbol{a}'} \alpha_{\boldsymbol{q}}(\boldsymbol{a}) = \sum_{\boldsymbol{a} \in A^r \colon \boldsymbol{a}_{[1,k]} = \boldsymbol{a}'} \sum_{\boldsymbol{a}'' \in A^k} \theta'(\boldsymbol{a}, \boldsymbol{a}'' \mid \boldsymbol{q}, \boldsymbol{q}_{[1,k]}),$$

while

$$\beta_{\boldsymbol{q}_{[1,k]}}(\boldsymbol{a}') = \sum_{\boldsymbol{a} \in A^r} \theta'(\boldsymbol{a}, \boldsymbol{a}' \mid \boldsymbol{q}, \boldsymbol{q}_{[1,k]}).$$

Therefore,

$$
\begin{aligned}
\mathrm{SD}(\alpha_{\boldsymbol{q},k}, \beta_{\boldsymbol{q}_{[1,k]}}) &= \frac{1}{2} \sum_{\boldsymbol{a}' \in A^k} \left| \alpha_{\boldsymbol{q},k}(\boldsymbol{a}') - \beta_{\boldsymbol{q}_{[1,k]}}(\boldsymbol{a}') \right| \\
&= \frac{1}{2} \sum_{\boldsymbol{a}' \in A^k} \left| \sum_{\boldsymbol{a} \in A^r \colon \boldsymbol{a}_{[1,k]} = \boldsymbol{a}'} \sum_{\boldsymbol{a}'' \in A^k} \theta'(\boldsymbol{a}, \boldsymbol{a}'' \mid \boldsymbol{q}, \boldsymbol{q}_{[1,k]}) - \sum_{\boldsymbol{a} \in A^r} \theta'(\boldsymbol{a}, \boldsymbol{a}' \mid \boldsymbol{q}, \boldsymbol{q}_{[1,k]}) \right| \\
&= \frac{1}{2} \sum_{\boldsymbol{a}' \in A^k} \left| \sum_{\boldsymbol{a} \in A^r \colon \boldsymbol{a}_{[1,k]} = \boldsymbol{a}'} \sum_{\boldsymbol{a}'' \in A^k \colon \boldsymbol{a}'' \neq \boldsymbol{a}'} \theta'(\boldsymbol{a}, \boldsymbol{a}'' \mid \boldsymbol{q}, \boldsymbol{q}_{[1,k]}) - \sum_{\boldsymbol{a} \in A^r \colon \boldsymbol{a}_{[1,k]} \neq \boldsymbol{a}'} \theta'(\boldsymbol{a}, \boldsymbol{a}' \mid \boldsymbol{q}, \boldsymbol{q}_{[1,k]}) \right| \\
&\leq \frac{1}{2} \sum_{\boldsymbol{a}' \in A^k} \left( \sum_{\boldsymbol{a} \in A^r \colon \boldsymbol{a}_{[1,k]} = \boldsymbol{a}'} \sum_{\boldsymbol{a}'' \in A^k \colon \boldsymbol{a}'' \neq \boldsymbol{a}'} \theta'(\boldsymbol{a}, \boldsymbol{a}'' \mid \boldsymbol{q}, \boldsymbol{q}_{[1,k]}) + \sum_{\boldsymbol{a} \in A^r \colon \boldsymbol{a}_{[1,k]} \neq \boldsymbol{a}'} \theta'(\boldsymbol{a}, \boldsymbol{a}' \mid \boldsymbol{q}, \boldsymbol{q}_{[1,k]}) \right) \\
&= \sum_{\boldsymbol{a} \in A^r, \boldsymbol{a}' \in A^k \colon \boldsymbol{a}_{[1,k]} \neq \boldsymbol{a}'} \theta'(\boldsymbol{a}, \boldsymbol{a}' \mid \boldsymbol{q}, \boldsymbol{q}_{[1,k]}) \\
&= \varepsilon_{\mathrm{cons}}(\boldsymbol{q}, k),
\end{aligned}
$$

as desired. □

Now we construct a strategy $\boldsymbol{\theta} = (\theta^{(1)}, \dots, \theta^{(r)})$ in $G$ as follows. Suppose that, just after the verifier has sent a question at the $k$th round, $k \in [r]$, the conversation between the verifier and the prover so far forms a sequence $(q_1, a_1, q_2, a_2, \dots, q_{k-1}, a_{k-1}, q_k)$, where each $q_j$, $j \in [k]$ is an element in $Q$, and each $a_j$, $j \in [k-1]$ is an element in $A$. Let $\boldsymbol{q}_{[1,k]} = (q_1, \dots, q_k)$ and $\boldsymbol{a}_{[1,k-1]} = (a_1, \dots, a_{k-1})$. The prover answers $a_k$ at this round with probability

$$
\theta^{(k)}(a_k \mid \boldsymbol{q}_{[1,k]}, \boldsymbol{a}_{[1,k-1]}) \overset{\text{def}}{=} \frac{\beta_{\boldsymbol{q}_{[1,k]}}(\boldsymbol{a}_{[1,k]})}{\sum_{a \in A} \beta_{\boldsymbol{q}_{[1,k]}}\big((\boldsymbol{a}_{[1,k-1]}, a)\big)}. \tag{3}
$$

If $k = 1$, the denominator of the right-hand side of Eq. (3) is treated as one. In case the denominator is zero, the value of $\theta^{(k)}(a_k \mid \boldsymbol{q}_{[1,k]}, \boldsymbol{a}_{[1,k-1]})$ is defined arbitrarily as long as it is nonnegative and its sum over $a_k$ is equal to one.

This strategy $\boldsymbol{\theta}$ can be interpreted as follows. At the first round, the prover in $G$ receives only one question, say $q_1 \in Q$. He simulates what the second prover would do in $G'$ when asked $q_1$, and reports the simulated answer $a_1 \in A$ to the verifier. At the $k$th round, $k \geq 2$, suppose that the conversation between the verifier and the prover until the $(k-1)$st round forms a sequence $(q_1, a_1, q_2, a_2, \dots, q_{k-1}, a_{k-1})$. Then the prover in $G$ when asked $q_k$ at the $k$th round simulates the behavior of the second prover in $G'$ with question $\boldsymbol{q}' = (q_1, \dots, q_k)$. If the simulated answer is of the form $\boldsymbol{a}' = (a_1, \dots, a_{k-1}, a_k)$ for some $a_k \in A$ (i.e., the first $k-1$ elements of the simulated answer are consistent with what he has answered so far), he responds with $a_k$ to the verifier. Otherwise (if the simulated answer contains an inconsist element), he discards it and retries the simulation until he obtains a consistent answer.

This $\boldsymbol{\theta}$ can be viewed as a *rounded strategy* in the terminology in Ref. [20]. It is noted, however, that $\boldsymbol{\theta}$ is different from the original rounded strategy used in Ref. [20] at least in the following two points:

(i) In the rounded strategy in Ref. [20], the prover uses the post-measurement quantum state. Since the concept of post-measurement states is not applicable to no-signaling strategies, the prover with strategy $\boldsymbol{\theta}$ restarts the simulation of the second prover in each round.

(ii) Because of (i), there is no guarantee that the prover in $G$ should obtain with high probability a simulation result that is consistent to what he has already answered. Therefore, in strategy $\boldsymbol{\theta}$, if the simulation result is inconsistent with the conversation so far, the prover restarts the simulation instead of aborting the game.

Hence, even if the no-signaling strategy $\theta'$ in $G'$ is in fact an entangled strategy, our construction gives a strategy $\boldsymbol{\theta}$ that is different from the rounded strategy used in Ref. [20]. The authors believe that this difference is essential when proving a bound for no-signaling strategies.

To bound $\varepsilon$ from below, we relate $\varepsilon$ to the winning probability of $\boldsymbol{\theta}$ in $G$ by hybrid argument. For each $k \in [r]$, let $h^{\langle k \rangle}$ be a family of probability distributions over $A^r$ indexed by each element in $Q^r$ defined by

$$h_{\boldsymbol{q}}^{\langle k \rangle}(\boldsymbol{a}) = \beta_{\boldsymbol{q}_{[1,k]}}(\boldsymbol{a}_{[1,k]}) \prod_{i=k+1}^{r} \theta^{(i)}(a_i \mid \boldsymbol{q}_{[1,i]}, \boldsymbol{a}_{[1,i-1]}).$$

Note that, in general, the family $h^{\langle k \rangle}$ of distributions for $k \geq 2$ is not induced from any valid strategy of the prover in $G$ because the prover has to know the first $k$ questions before answering to the first question. For each $k \in [r]$, let $p_k$ be the probability that this "imaginal strategy" $h^{\langle k \rangle}$ wins in $G$, i.e., let $p_k$ be such that

$$p_k = \sum_{\boldsymbol{q} \in Q^r} \pi(\boldsymbol{q}) \sum_{\boldsymbol{a} \in A^r} h_{\boldsymbol{q}}^{\langle k \rangle}(\boldsymbol{a}) R(\boldsymbol{a} \mid \boldsymbol{q}).$$

Note that $h^{\langle 1 \rangle}$ is identical to the family of probability distributions induced by the strategy $\boldsymbol{\theta}$, and thus, $p_1 \leq w(G)$. Also note that $h^{\langle r \rangle}$ is identical to the family of probability distributions induced by the behaviour of the second prover in $G'$ when asked a question in $Q^r$, and thus, $p_r \geq 1 - \varepsilon(r)$.

**Claim 2.** *For any $\boldsymbol{q} \in Q^r$ and any $k$, $2 \leq k \leq r$,*

$$\mathrm{SD}\big(h_{\boldsymbol{q}}^{\langle k-1 \rangle}, h_{\boldsymbol{q}}^{\langle k \rangle}\big) \leq \varepsilon_{\mathrm{cons}}(\boldsymbol{q}, k-1) + \varepsilon_{\mathrm{cons}}(\boldsymbol{q}, k).$$

*Proof.* By definition, for any $k$, $2 \leq k \leq r$, and for any $\boldsymbol{q} \in Q^r$ and $\boldsymbol{a} = (a_1, \ldots, a_r) \in A^{k-1}$,

$$\begin{aligned} &h_{\boldsymbol{q}}^{\langle k-1 \rangle}(\boldsymbol{a}) - h_{\boldsymbol{q}}^{\langle k \rangle}(\boldsymbol{a}) \\ &\qquad = \big(\beta_{\boldsymbol{q}_{[1,k-1]}}(\boldsymbol{a}_{[1,k-1]})\theta^{(k)}(a_k \mid \boldsymbol{q}_{[1,k]}, \boldsymbol{a}_{[1,k-1]}) - \beta_{\boldsymbol{q}_{[1,k]}}(\boldsymbol{a}_{[1,k]})\big) \prod_{i=k+1}^{r} \theta^{(i)}(a_i \mid \boldsymbol{q}_{[1,i]}, \boldsymbol{a}_{[1,i-1]}). \end{aligned}$$

This implies that, for any $k$, $2 \leq k \leq r$, and for any $\boldsymbol{q} \in Q^r$ and $\boldsymbol{a}' \in A^{k-1}$,

$$\begin{aligned} &\sum_{a_k, \ldots, a_r \in A} \Big| h_{\boldsymbol{q}}^{\langle k-1 \rangle}\big((\boldsymbol{a}', a_k, \ldots, a_r)\big) - h_{\boldsymbol{q}}^{\langle k \rangle}\big((\boldsymbol{a}', a_k, \ldots, a_r)\big) \Big| \\ &\quad = \sum_{a_k \in A} \big| \beta_{\boldsymbol{q}_{[1,k-1]}}(\boldsymbol{a}')\theta^{(k)}(a_k \mid \boldsymbol{q}_{[1,k]}, \boldsymbol{a}') - \beta_{\boldsymbol{q}_{[1,k]}}\big((\boldsymbol{a}', a_k)\big)(\boldsymbol{a}') \big| \\ &\quad = \sum_{a_k \in A} \left| \beta_{\boldsymbol{q}_{[1,k-1]}}(\boldsymbol{a}')\theta^{(k)}(a_k \mid \boldsymbol{q}_{[1,k]}, \boldsymbol{a}') - \theta^{(k)}(a_k \mid \boldsymbol{q}_{[1,k]}, \boldsymbol{a}') \sum_{a \in A} \beta_{\boldsymbol{q}_{[1,k]}}\big((\boldsymbol{a}', a)\big) \right| \\ &\quad = \left( \sum_{a_k \in A} \theta^{(k)}(a_k \mid \boldsymbol{q}_{[1,k]}, \boldsymbol{a}') \right) \left| \beta_{\boldsymbol{q}_{[1,k-1]}}(\boldsymbol{a}') - \sum_{a \in A} \beta_{\boldsymbol{q}_{[1,k]}}\big((\boldsymbol{a}', a)\big) \right| \\ &\quad = \left| \beta_{\boldsymbol{q}_{[1,k-1]}}(\boldsymbol{a}') - \sum_{a \in A} \beta_{\boldsymbol{q}_{[1,k]}}\big((\boldsymbol{a}', a)\big) \right|, \end{aligned}$$

where the second equality follows from the fact that the relation

$$\beta_{\boldsymbol{q}_{[1,k]}}(\boldsymbol{a}_{[1,k]}) = \theta^{(k)}(a_k \mid \boldsymbol{q}_{[1,k]}, \boldsymbol{a}_{[1,k-1]}) \sum_{a \in A} \beta_{\boldsymbol{q}_{[1,k]}}\big((\boldsymbol{a}_{[1,k-1]}, a)\big)$$

holds no matter whether $\sum_{a \in A} \beta_{\boldsymbol{q}_{[1,k]}}\big((\boldsymbol{a}_{[1,k-1]}, a)\big) > 0$ or not. Hence,

$$\mathrm{SD}\big(h_{\boldsymbol{q}}^{\langle k-1\rangle}, h_{\boldsymbol{q}}^{\langle k\rangle}\big) = \frac{1}{2} \sum_{\boldsymbol{a} \in A^r} \big|h_{\boldsymbol{q}}^{\langle k-1\rangle}(\boldsymbol{a}) - h_{\boldsymbol{q}}^{\langle k\rangle}(\boldsymbol{a})\big| = \frac{1}{2} \sum_{\boldsymbol{a}' \in A^{k-1}} \left|\beta_{\boldsymbol{q}_{[1,k-1]}}(\boldsymbol{a}') - \sum_{a \in A} \beta_{\boldsymbol{q}_{[1,k]}}\big((\boldsymbol{a}', a)\big)\right|.$$

It follows from the triangle inequality that

$$\mathrm{SD}\big(h_{\boldsymbol{q}}^{\langle k-1\rangle}, h_{\boldsymbol{q}}^{\langle k\rangle}\big) \leq \frac{1}{2} \sum_{\boldsymbol{a}' \in A^{k-1}} \left|\beta_{\boldsymbol{q}_{[1,k-1]}}(\boldsymbol{a}') - \alpha_{\boldsymbol{q},k-1}(\boldsymbol{a}')\right| + \frac{1}{2} \sum_{\boldsymbol{a}' \in A^{k-1}} \left|\alpha_{\boldsymbol{q},k-1}(\boldsymbol{a}') - \sum_{a \in A} \beta_{\boldsymbol{q}_{[1,k]}}\big((\boldsymbol{a}', a)\big)\right|. \quad (4)$$

The first term of the right-hand side of the inequality (4) is exactly $\mathrm{SD}(\alpha_{\boldsymbol{q},k-1}, \beta_{\boldsymbol{q}_{[1,k-1]}})$, which is at most $\varepsilon_{\mathrm{cons}}(\boldsymbol{q}, k-1)$ by Claim 1. The second term of the right-hand side of the inequality (4) is bounded from above as

$$\begin{aligned}
\frac{1}{2} \sum_{\boldsymbol{a}' \in A^{k-1}} \left|\alpha_{\boldsymbol{q},k-1}(\boldsymbol{a}') - \sum_{a \in A} \beta_{\boldsymbol{q}_{[1,k]}}\big((\boldsymbol{a}', a)\big)\right| &= \frac{1}{2} \sum_{\boldsymbol{a}' \in A^{k-1}} \left|\sum_{a \in A} \alpha_{\boldsymbol{q},k}\big((\boldsymbol{a}', a)\big) - \sum_{a \in A} \beta_{\boldsymbol{q}_{[1,k]}}\big((\boldsymbol{a}', a)\big)\right| \\
&\leq \frac{1}{2} \sum_{\boldsymbol{a}'' \in A^{k}} \left|\alpha_{\boldsymbol{q},k}(\boldsymbol{a}'') - \beta_{\boldsymbol{q}_{[1,k]}}(\boldsymbol{a}'')\right| \\
&= \mathrm{SD}(\alpha_{\boldsymbol{q},k}, \beta_{\boldsymbol{q}_{[1,k]}}),
\end{aligned}$$

which is at most $\varepsilon_{\mathrm{cons}}(\boldsymbol{q}, k)$ again by Claim 1.

Hence, $\mathrm{SD}\big(h_{\boldsymbol{q}}^{\langle k-1\rangle}, h_{\boldsymbol{q}}^{\langle k\rangle}\big) \leq \varepsilon_{\mathrm{cons}}(\boldsymbol{q}, k-1) + \varepsilon_{\mathrm{cons}}(\boldsymbol{q}, k)$, as claimed. □

Claim 2 combined with the triangle inequality implies

$$\mathrm{SD}\big(h_{\boldsymbol{q}}^{\langle 1\rangle}, h_{\boldsymbol{q}}^{\langle r\rangle}\big) \leq \sum_{k=2}^{r} \mathrm{SD}\big(h_{\boldsymbol{q}}^{\langle k-1\rangle}, h_{\boldsymbol{q}}^{\langle k\rangle}\big) \leq \sum_{k=2}^{r} (\varepsilon_{\mathrm{cons}}(\boldsymbol{q}, k-1) + \varepsilon_{\mathrm{cons}}(\boldsymbol{q}, k)) \leq 2 \sum_{k=1}^{r} \varepsilon_{\mathrm{cons}}(\boldsymbol{q}, k) = 2r\varepsilon_{\mathrm{cons}}(\boldsymbol{q}).$$

Hence,

$$\begin{aligned}
p_r - p_1 &= \sum_{\boldsymbol{q} \in Q^r} \pi(\boldsymbol{q}) \sum_{\boldsymbol{a} \in A^r} h_{\boldsymbol{q}}^{\langle r\rangle}(\boldsymbol{a}) R(\boldsymbol{a} \mid \boldsymbol{q}) - \sum_{\boldsymbol{q} \in Q^r} \pi(\boldsymbol{q}) \sum_{\boldsymbol{a} \in A^r} h_{\boldsymbol{q}}^{\langle 1\rangle}(\boldsymbol{a}) R(\boldsymbol{a} \mid \boldsymbol{q}) \\
&= \sum_{\boldsymbol{q} \in Q^r} \pi(\boldsymbol{q}) \sum_{\boldsymbol{a} \in A^r} \big(h_{\boldsymbol{q}}^{\langle r\rangle}(\boldsymbol{a}) R(\boldsymbol{a} \mid \boldsymbol{q}) - h_{\boldsymbol{q}}^{\langle 1\rangle}(\boldsymbol{a}) R(\boldsymbol{a} \mid \boldsymbol{q})\big) \\
&\leq \sum_{\boldsymbol{q} \in Q^r} \pi(\boldsymbol{q}) \mathrm{SD}\big(h_{\boldsymbol{q}}^{\langle 1\rangle}, h_{\boldsymbol{q}}^{\langle r\rangle}\big) \\
&\leq 2r \sum_{\boldsymbol{q} \in Q^r} \pi(\boldsymbol{q}) \varepsilon_{\mathrm{cons}}(\boldsymbol{q}) \\
&= 2r\varepsilon_{\mathrm{cons}},
\end{aligned}$$

where the first inequality is due to the basic property of the statistical difference. Thus,

$$p_r \leq p_1 + 2r\varepsilon_{\mathrm{cons}} \leq p_1 + 2r\varepsilon.$$

Since $p_1 \leq w(G)$ and $p_r \geq 1 - \varepsilon(r)$, it follows from the inequality (2) that

$$r\varepsilon \geq \varepsilon(r) \geq 1 - p_r \geq 1 - p_1 - 2r\varepsilon \geq 1 - w(G) - 2r\varepsilon,$$

or $\varepsilon \geq \frac{1-w(G)}{3r}$, as desired.

# 4 Oracularization with a dummy question

This section defines oracularization with a dummy question and proves Theorem 2 which states that NEXP has a two-prover one-round interactive proof system which is sound against entangled provers in a weak sense. Corollary 3, the scaled-down version of Theorem 2, is also proved.

## 4.1 Definition of the transformation

Theorem 2 and Corollary 3 are proved by transforming non-adaptive three-query PCP games to two-prover one-round games which are sound in a weak sense against dishonest entangled provers.

Let $G = (\pi, V)$ be a nonadaptive three-query PCP game. *Oracularization with a dummy question* of $G$ gives a two-prover one-round game $G' = ([Q]^3, [Q]^2, A^3, A^2, R', \pi')$, where the predicate $R'$ and the probability distribution $\pi'$ are specified as follows.

The verifier $V'$ in $G'$ chooses two triples $(q_1, q_2, q_3), (\tilde{q}_1, \tilde{q}_2, \tilde{q}_3) \in [Q]^3$ according to the distribution $\pi$. Then he chooses $q \in \{q_1, q_2, q_3\}$ and $\tilde{q} \in \{\tilde{q}_1, \tilde{q}_2, \tilde{q}_3\}$ uniformly and independently at random. He decides $r \in \{1, 2\}$ according to the following rule: $r = 1$ if $q < \tilde{q}$, $r = 2$ if $q > \tilde{q}$, and choose $r \in \{1, 2\}$ uniformly at random if $q = \tilde{q}$. He sends questions $(q_1, q_2, q_3)$ to the first prover. If $r = 1$, then he sends $(q, \tilde{q})$ to the second prover, and otherwise he sends $(\tilde{q}, q)$. This specifies the probability distribution $\pi'$ over the set $[Q]^3 \times [Q]^2$. Upon receiving answers $(a_1, a_2, a_3) \in A^3$ from the first prover and $(a'_1, a'_2) \in A^2$ from the second prover, $V'$ performs the following two tests:

(SIMULATION TEST) $V'$ verifies that $R(a_1, a_2, a_3 \mid q_1, q_2, q_3) = 1$,

(CONSISTENCY TEST) $V'$ verifies that $a'_r = a_j$ when $q = q_j$.

The provers win if and only if they pass both tests. This specifies the predicate $R' \colon [Q]^3 \times [Q]^2 \times A^3 \times A^2 \to \{0, 1\}$.

This transformation is different from the usual oracularization process of nonadaptive three-query PCP games explained in Subsection 2.3 in that $V'$ also sends a *dummy question* $\tilde{q}$ to the second prover without letting the prover know which question is "real."

Note that without adding a dummy question, entangled provers could win perfectly in some games even if the unentangled provers cannot win perfectly, as noted in the introduction. On the other hand, we do not know whether the oracularization technique is necessary to bound the entangled value. That is, if we just add a dummy question to each prover without first applying the oracularization technique to the protocol, we do not know whether it is possible to bound the entangled value.

The main lemma on oracularization with a dummy question is as follows.

**Lemma 8.** *There exists a constant $c > 0$ such that for any nonadaptive three-query PCP game $G = (Q, A, R, \pi)$,*

$$w(G) \le w_{\text{unent}}(G') \le 1 - \frac{1 - w(G)}{3},$$

$$w_{\text{com}}(G') \le 1 - \frac{c(1 - w(G))^2}{Q^2},$$

*where $G'$ is the two-prover one-round game constructed by oracularization with a dummy question from $G$.*

## 4.2 Proofs of Theorem 2 and Corollary 3

Here we prove Theorem 2 and Corollary 3 assuming Lemma 8. We need the following form of the PCP theorem based on the result of Khanna, Sudan, Trevisan and Williamson [22].

**Theorem 9.** *(i) There exists a constant $0 < s < 1$ such that every language in* $\mathrm{NEXP}$ *has a nonadaptive three-query PCP system of perfect completeness with soundness error at most $s$ and free-bit complexity $\log_2 3$ (that is, the verifier accepts at most three out of the $2^3 = 8$ possible results of reading).*

*(ii) There exists a constant $0 < s < 1$ such that every language in* $\mathrm{NP}$ *has a nonadaptive three-query PCP system of perfect completeness with soundness error at most $s$ and free-bit complexity $\log_2 3$ where the verifier uses $O(\log n)$ randomness.*

*Proof.* Corollary 5.13 and Theorem 5.14 of Khanna, Sudan, Trevisan and Williamson [22] implies that given an instance of 1-IN-3 3SAT, it is NP-complete to distinguish whether it is satisfiable or no assignments satisfy more than a constant fraction of clauses. Part (ii) follows by restating this result in terms of PCP systems. Part (i) follows since the reduction in Ref. [22] is a local transformation and scales up to NEXP-completeness of the succinct version of the problem. □

*Proof of Theorem 2.* Apply oracularization with a dummy question to the PCP system in Theorem 9 (i) to obtain a two-prover one-round interactive proof system. By Lemma 8, this two-prover one-round interactive proof system satisfies properties (ii), (iii) and (iv). In this two-prover one-round interactive proof system, the first prover responds with three bits and the second prover responds with two bits. However, since the PCP system has free-bit complexity $\log_2 3$, there are at most three possibilities for the answer of the first prover in the honest case, and therefore the answer of the first prover can be encoded in one trit. □

*Proof of Corollary 3.* Membership to NP is clear since the problem is at most as hard as deciding whether or not $w_{\mathrm{unent}}(G) = 1$ ignoring the promise.

NP-hardness follows by applying Lemma 8 to the PCP system in Theorem 9 (ii) in the same way as the proof of Theorem 2. □

### 4.3 Proof of Lemma 8

$w_{\mathrm{unent}}(G') \le 1 - (1 - w(G))/3$ follows from Theorem 5. $w_{\mathrm{unent}}(G') \ge w(G)$ is easy: let $\Pi \in A^Q$ be a string which makes $V$ accept with probability $w(G)$. The provers in $G'$ achieve the same winning probability $w(G)$ if the first prover answers $(\Pi[q_1], \Pi[q_2], \Pi[q_3])$ upon questions $(q_1, q_2, q_3)$, and the second prover answers $(\Pi[q'_1], \Pi[q'_2])$ upon questions $(q'_1, q'_2)$.

In the rest of the section, we shall prove $w_{\mathrm{com}}(G') \le 1 - c(1 - w(G))^2/Q^2$, where $c = 1/(1 + 15\sqrt{2})^2$.

Without loss of generality, we assume that all positions $q \in [Q]$ are queried with nonzero probability; otherwise drop the positions which are never queried. We define a probability distribution $\pi(q)$ over $q \in [Q]$ as

$$\pi(q) = \frac{1}{3} \sum_{1 \le i \le 3} \sum_{\substack{1 \le q_1 < q_2 < q_3 \le Q \\ q_i = q}} \pi(q_1, q_2, q_3).$$

If one chooses questions $(q_1, q_2, q_3) \in [Q]^3$ according to $\pi(q_1, q_2, q_3)$ and chooses one of $q_1, q_2, q_3$ uniformly at random, then $q$ is chosen with probability $\pi(q)$. Also without loss of generality, we assume that the $Q$ positions are labelled in the descending order of their marginal probabilities: $\pi(1) \ge \pi(2) \ge \cdots \ge \pi(Q) > 0$.

In the analysis of the soundness of the protocol, we use the following lemma.

**Lemma 10.** *Let $\boldsymbol{M} = \{M_a\}_{a \in A}$ and $\boldsymbol{N} = \{N_a\}_{a \in A}$ be mutually commuting POVMs on a Hilbert space $\mathcal{H}$. Let $|\varphi\rangle \in \mathcal{H}$ be a state, and define two states $|\psi\rangle, |\xi\rangle \in \mathbb{C}^A \otimes \mathcal{H}$ by*

$$|\psi\rangle = \sum_{a \in A} |a\rangle \otimes \sqrt{M_a}\, |\varphi\rangle,$$

$$|\xi\rangle = \sum_{a \in A} |a\rangle \otimes \sqrt{N_a}\, |\varphi\rangle.$$

*Let $p$ be the probability with which joint measurement of $\boldsymbol{M}$ and $\boldsymbol{N}$ on $|\varphi\rangle$ gives different results:*

$$p = 1 - \sum_{a \in A} \langle\varphi|M_a N_a|\varphi\rangle.$$

*Then,*

$$D^2(|\psi\rangle, |\xi\rangle) \le 2(1 - \langle\psi|\xi\rangle) \le 2p.$$

*Proof.* The first inequality is proved by Fact 1:

$$\begin{aligned} D^2(|\psi\rangle, |\xi\rangle) &= 1 - (\langle\psi|\xi\rangle)^2 \\ &= (1 + \langle\psi|\xi\rangle)(1 - \langle\psi|\xi\rangle) \\ &\le 2(1 - \langle\psi|\xi\rangle). \end{aligned}$$

The second inequality is proved as:

$$\begin{aligned} 1 - \langle\psi|\xi\rangle &= 1 - \sum_{a \in A} \langle\varphi|\sqrt{M_a}\sqrt{N_a}\,|\varphi\rangle \\ &= 1 - \sum_{a \in A} \langle\varphi|\sqrt{M_a N_a}\,|\varphi\rangle \\ &\le 1 - \sum_{a \in A} \langle\varphi|M_a N_a|\varphi\rangle = p. \end{aligned}$$

□

Consider an arbitrary commuting-operator strategy in $G'$, and let $1 - \varepsilon$ be its winning probability. Without loss of generality, we assume that the state shared by the provers is pure and that every measurement performed by the provers is projective.

As a preprocessing, we convert this strategy to a commuting-operator strategy where the second prover answers two identical answers whenever asked two identical questions. To achieve this, the new provers share an additional EPR pair $(|00\rangle + |11\rangle)/\sqrt{2}$. The new second prover acts exactly the same way as the original second prover if his two questions are distinct. If his two questions are identical, then he first simulates the original second prover to obtain answers $(a'_1, a'_2) \in A^2$, uses the added EPR pair to simulate a classical coin flip to choose $a' \in \{a'_1, a'_2\}$ uniformly at random, and answers $(a', a')$. This preprocessing does not change the winning probability, the shared state is still pure, and every measurement is still projective.

Let $|\Psi\rangle \in \mathcal{P}$ be a quantum state shared by the provers. For $1 \le q_1 < q_2 < q_3 \le Q$, let $\boldsymbol{M}_{q_1q_2q_3} = \{M^{a_1a_2a_3}_{q_1q_2q_3}\}_{a_1,a_2,a_3 \in A}$ be the PVM measured by the first prover upon the questions $(q_1, q_2, q_3)$. For $1 \le q'_1 \le q'_2 \le Q$, let $\boldsymbol{N}_{q'_1q'_2} = \{N^{a'_1a'_2}_{q'_1q'_2}\}_{a'_1,a'_2 \in A}$ be the PVM measured by the second prover upon the questions $q'_1, q'_2$. We use the notation $\boldsymbol{M}_{q_1q_2q_3} = \{M^{a_1a_2a_3}_{q_1q_2q_3}\}_{a_1,a_2,a_3 \in A}$ for three distinct questions $q_1, q_2, q_3 \in [Q]$ even when they do not satisfy $q_1 < q_2 < q_3$. In this case, we define $M^{a_1a_2a_3}_{q_1q_2q_3} = M^{a_{\sigma(1)}a_{\sigma(2)}a_{\sigma(3)}}_{q_{\sigma(1)}q_{\sigma(2)}q_{\sigma(3)}}$, where $\sigma$ is a permutation on $\{1, 2, 3\}$ such that $q_{\sigma(1)} < q_{\sigma(2)} < q_{\sigma(3)}$. Similarly, if $q'_1 > q'_2$, let $\boldsymbol{N}_{q'_1q'_2} = \{N^{a'_1a'_2}_{q'_1q'_2}\}_{a'_1,a'_2 \in A}$ be a PVM defined by $N^{a'_1a'_2}_{q'_1q'_2} = N^{a'_2a'_1}_{q'_2q'_1}$. By the preprocessing above, $N^{a'_1a'_2}_{q'_1q'_2} = N^{a'_2a'_1}_{q'_2q'_1}$ no matter whether $q'_1 = q'_2$ or not.

By the definition of no-signaling strategies, $\boldsymbol{M}_{q_1q_2q_3}$ and $\boldsymbol{N}_{q\tilde{q}}$ mutually commute. Then, under the condition that the verifier sends $q_1, q_2, q_3$ to the first prover and $q, \tilde{q}$ to the second prover, the probability with which the first prover returns answers $a_1, a_2, a_3$ for $q_1, q_2, q_3$, respectively, and the second prover returns answers $b$ for $q$ and $\tilde{a}$ for $\tilde{q}$ is given by $\langle\Psi|M^{a_1a_2a_3}_{q_1q_2q_3}N^{a\tilde{a}}_{q\tilde{q}}|\Psi\rangle$. We denote the summation of PVM operators over all answers like $M^{a_1}_{q_1|q_1q_2q_3} = \sum_{a_2,a_3 \in A} M^{a_1a_2a_3}_{q_1q_2q_3}$ or $N^{a}_{q|q\tilde{q}} = \sum_{\tilde{a} \in A} N^{a\tilde{a}}_{q\tilde{q}}$. For $q \in [Q]$, we define POVMs $\bar{\boldsymbol{M}}_q = \{\bar{M}^a_q\}_{a \in A}$ and

$\bar{\boldsymbol{N}}_q = \{\bar{N}_q^a\}_{a\in A}$ by

$$\bar{M}_q^a = \frac{1}{\pi(q)} \sum_{1\le i\le 3} \sum_{\substack{1\le q_1<q_2<q_3\le Q\\ q_i=q}} \pi(q_1,q_2,q_3) M^a_{q|q_1q_2q_3},$$
$$\bar{N}_q^a = \sum_{\tilde{q}\in[Q]} \pi(\tilde{q}) N^a_{q|q\tilde{q}}.$$

Then the winning probability of this strategy is

$$\begin{aligned}1-\varepsilon &= \frac{1}{9} \sum_{1\le q_1<q_2<q_3\le Q} \sum_{1\le i\le 3} \sum_{1\le \tilde{q}_1<\tilde{q}_2<\tilde{q}_3\le Q} \sum_{\tilde{q}\in\{\tilde{q}_1,\tilde{q}_2,\tilde{q}_3\}} \pi(q_1,q_2,q_3)\pi(\tilde{q}_1,\tilde{q}_2,\tilde{q}_3) \times \\ &\qquad \sum_{a_1,a_2,a_3\in A} \langle\Psi| M^{a_1a_2a_3}_{q_1q_2q_3} N^{a_i}_{q_i|q_i\tilde{q}} |\Psi\rangle R(a_1,a_2,a_3 \mid q_1,q_2,q_3) \\ &= \frac{1}{3} \sum_{1\le q_1<q_2<q_3\le Q} \sum_{1\le i\le 3} \pi(q_1,q_2,q_3) \sum_{a_1,a_2,a_3\in A} \langle\Psi| M^{a_1a_2a_3}_{q_1q_2q_3} \bar{N}^{a_i}_{q_i} |\Psi\rangle R(a_1,a_2,a_3 \mid q_1,q_2,q_3).\end{aligned}$$

This strategy passes the CONSISTENCY TEST with probability

$$1-\varepsilon_{\text{cons}} = \sum_{q\in[Q]} \pi(q) \sum_{a\in A} \langle\Psi|\bar{M}_q^a \bar{N}_q^a|\Psi\rangle,$$

and the SIMULATION TEST with probability

$$1-\varepsilon_{\text{sim}} = \sum_{1\le q_1<q_2<q_3\le Q} \pi(q_1,q_2,q_3) \sum_{a_1,a_2,a_3\in A} \langle\Psi| M^{a_1a_2a_3}_{q_1q_2q_3} |\Psi\rangle R(a_1,a_2,a_3 \mid q_1,q_2,q_3).$$

Note that $\varepsilon \ge \varepsilon_{\text{cons}}$ and $\varepsilon \ge \varepsilon_{\text{sim}}$.

For simplicity of notation, let $X_q^a = \sqrt{\bar{M}_q^a}$ and $Y_q^a = \sqrt{\bar{N}_q^a}$. We define a strategy $\theta$ in the game $G$ by

$$\theta(\Pi) = \| X_Q^{\Pi[Q]} X_{Q-1}^{\Pi[Q-1]} \cdots X_1^{\Pi[1]} |\Psi\rangle \|^2.$$

This distribution can be explained as follows. The prover in $G$ prepares the quantum state $|\Psi\rangle$, and measures the state using POVM $\{\bar{M}_1^{a_1}\}_{a_1\in A}$ to obtain a classical answer $a_1 \in A$ and a post-measurement state $X_1^{a_1}|\Psi\rangle$. Next, he measures this post-measurement state using POVM $\{\bar{M}_2^{a_2}\}_{a_2\in A}$ to obtain $a_2 \in A$ and a post-measurement state $X_2^{a_2}X_1^{a_1}|\Psi\rangle$. He repeats this for all the $Q$ questions. Note that the order of the questions is significant. This strategy naturally induces

$$\theta(a_1,a_2,a_3 \mid q_1,q_2,q_3) = \sum_{\substack{\Pi\in A^Q\\ \Pi[q_i]=a_i\ (i=1,2,3)}} \theta(\Pi).$$

By assumption,

$$\sum_{1\le q_1<q_2<q_3\le Q} \pi(q_1,q_2,q_3) \sum_{a_1,a_2,a_3\in A} \theta(a_1,a_2,a_3 \mid q_1,q_2,q_3) R(a_1,a_2,a_3 \mid q_1,q_2,q_3) \le w(G). \tag{5}$$

Note the similarity of this construction to the "rounding" by Kempe, Kobayashi, Matsumoto, Toner and Vidick [20]. Following Ref. [20], we shall bound the winning probability $1-\varepsilon$ of the strategy in $G'$ by proving that (a) the measurements $\bar{\boldsymbol{M}}_q$ and $\bar{\boldsymbol{N}}_q$ are close, and (b) the operators $X_q^b$ are pairwise almost commuting, which roughly correspond to Claim 14 of Ref. [20]. The fact that a dummy question is chosen independently from the rest of the questions is important when proving these properties. Though the rest of the proof is similar to that of Ref. [20], our proof diverges from this basic idea superficially to obtain a little better bound.

Let

$$\begin{aligned} d_1(q) &= D\left(\sum_{a\in A}|a\rangle\otimes X_q^a|\Psi\rangle, \sum_{a\in A}|a\rangle\otimes Y_q^a|\Psi\rangle\right), \quad q\in[Q], \\ d_2(q\mid\tilde{q}) &= D\left(\sum_{a\in A}|a\rangle\otimes X_q^a|\Psi\rangle, \sum_{a\in A}|a\rangle\otimes N_{q|q\tilde{q}}^a|\Psi\rangle\right), \quad q,\tilde{q}\in[Q], \\ d_3(q\mid q_1q_2q_3) &= D\left(\sum_{a\in A}|a\rangle\otimes M_{q|q_1q_2q_3}^a|\Psi\rangle, \sum_{a\in A}|a\rangle\otimes Y_q^a|\Psi\rangle\right), \quad 1\le q_1<q_2<q_3\le Q,\ q\in\{q_1,q_2,q_3\}, \\ d_4(q_1,q_2) &= D\left(\sum_{a_1,a_2\in A}|a_1\rangle|a_2\rangle\otimes X_{q_2}^{a_2}X_{q_1}^{a_1}|\Psi\rangle, \sum_{a_1,a_2\in A}|a_1\rangle|a_2\rangle\otimes X_{q_1}^{a_1}X_{q_2}^{a_2}|\Psi\rangle\right), \quad q_1,q_2\in[Q]. \end{aligned}$$

For notational convenience, we denote $(d_1(q))^2$ as $d_1^2(q)$, $d_2(q\mid q'))^2$ as $d_2^2(q\mid q')$, and so on.

**Claim 3.**

$$\mathbf{E}_q[d_1^2(q)] \le 2\varepsilon_{\mathrm{cons}}, \tag{6}$$

$$\mathbf{E}_{q,\tilde{q}}[d_2^2(q\mid\tilde{q})] \le 2\varepsilon_{\mathrm{cons}}, \tag{7}$$

$$\mathbf{E}_{(q_1,q_2,q_3)}\,\mathbf{E}_i[d_3^2(q_i\mid q_1q_2q_3)] \le 2\varepsilon_{\mathrm{cons}}, \tag{8}$$

$$\mathbf{E}_{q_1,q_2}[d_4^2(q_1,q_2)] \le 32\varepsilon_{\mathrm{cons}}, \tag{9}$$

*where $\mathbf{E}_q[]$ denotes the average over $q\in[Q]$ weighted by $\pi(q)$, $\mathbf{E}_{q_1,q_2}[]$ denotes the average over $q_1,q_2\in[Q]$ weighted by $\pi(q_1)\pi(q_2)$, $\mathbf{E}_{(q_1,q_2,q_3)}[]$ denotes the average over $q_1,q_2,q_3\in[Q]$ weighted by $\pi(q_1,q_2,q_3)$, and $\mathbf{E}_i[]$ denotes the average over $i\in\{1,2,3\}$ weighted uniformly.*

*Proof.* Eqs. (6), (7) and (8) are proved in a similar way by using Lemma 10. For Eq. (6), by Lemma 10,

$$\begin{aligned} d_1^2(q) &= D^2\left(\sum_{a\in A}|a\rangle\otimes X_q^a|\Psi\rangle, \sum_{a\in A}|a\rangle\otimes Y_q^a|\Psi\rangle\right) \\ &\le 2\left(1-\sum_{a\in A}\langle\Psi|\bar{M}_q^a\bar{N}_q^a|\Psi\rangle\right), \end{aligned}$$

which implies

$$\mathbf{E}_q[d_1^2(q)] \le 2\,\mathbf{E}_q\left[1-\sum_{a\in A}\langle\Psi|\bar{M}_q^a\bar{N}_q^a|\Psi\rangle\right] = 2\varepsilon_{\mathrm{cons}}.$$

For Eq. (7), by Lemma 10,

$$d_2^2(q\mid\tilde{q}) \le 2\left(1-\sum_{a\in A}\langle\Psi|\bar{M}_q^a N_{q|q\tilde{q}}^a|\Psi\rangle\right),$$

which implies

$$\mathbf{E}_{q,\tilde{q}}[d_2^2(q\mid\tilde{q})] \le 2\,\mathbf{E}_{q,\tilde{q}}\left[1-\sum_{a\in A}\langle\Psi|\bar{M}_q^a N_{q|q\tilde{q}}^a|\Psi\rangle\right] = 2\varepsilon_{\mathrm{cons}}.$$

For Eq. (8), again by Lemma 10,

$$d_3^2(q \mid q_1q_2q_3) \le 2\left(1 - \sum_{a\in A}\langle\Psi|M^a_{q|q_1q_2q_3}\bar{N}^a_q|\Psi\rangle\right),$$

which implies

$$\mathbf{E}_{(q_1,q_2,q_3)}\,\mathbf{E}_i[d_3^2(q_i \mid q_1q_2q_3)] \le 2\,\mathbf{E}_{(q_1,q_2,q_3)}\,\mathbf{E}_i\left[1 - \sum_{a\in A}\langle\Psi|M^a_{q_i|q_1q_2q_3}\bar{N}^a_{q_i}|\Psi\rangle\right] = 2\varepsilon_{\mathrm{cons}}.$$

For Eq. (9),

$$D\left(\sum_{a_1,a_2\in A}|a_1\rangle|a_2\rangle\otimes X^{a_2}_{q_2}X^{a_1}_{q_1}|\Psi\rangle, \sum_{a_1,a_2\in A}|a_1\rangle|a_2\rangle\otimes X^{a_2}_{q_2}N^{a_1}_{q_1|q_1q_2}|\Psi\rangle\right) \le d_2(q_1 \mid q_2), \tag{10a}$$

$$X^{a_2}_{q_2}N^{a_1}_{q_1|q_1q_2} = N^{a_1}_{q_1|q_1q_2}X^{a_2}_{q_2}, \tag{10b}$$

$$D\left(\sum_{a_1,a_2\in A}|a_1\rangle|a_2\rangle\otimes N^{a_1}_{q_1|q_1q_2}X^{a_2}_{q_2}|\Psi\rangle, \sum_{a_1,a_2\in A}|a_1\rangle|a_2\rangle\otimes N^{a_1}_{q_1|q_1q_2}N^{a_2}_{q_2|q_1q_2}|\Psi\rangle\right) \le d_2(q_2 \mid q_1), \tag{10c}$$

$$N^{a_1}_{q_1|q_1q_2}N^{a_2}_{q_1|q_1q_2} = N^{a_1a_2}_{q_1q_2} = N^{a_2}_{q_2|q_1q_2}N^{a_1}_{q_1|q_1q_2}, \tag{10d}$$

$$D\left(\sum_{a_1,a_2\in A}|a_1\rangle|a_2\rangle\otimes N^{a_2}_{q_2|q_1q_2}N^{a_1}_{q_1|q_1q_2}|\Psi\rangle, \sum_{a_1,a_2\in A}|a_1\rangle|a_2\rangle\otimes N^{a_2}_{q_2|q_1q_2}X^{a_1}_{q_1}|\Psi\rangle\right) \le d_2(q_1 \mid q_2), \tag{10e}$$

$$N^{a_2}_{q_2|q_1q_2}X^{a_1}_{q_1} = X^{a_1}_{q_1}N^{a_2}_{q_2|q_1q_2}, \tag{10f}$$

$$D\left(\sum_{a_1,a_2\in A}|a_1\rangle|a_2\rangle\otimes X^{a_1}_{q_1}N^{a_2}_{q_2|q_1q_2}|\Psi\rangle, \sum_{a_1,a_2\in A}|a_1\rangle|a_2\rangle\otimes X^{a_1}_{q_1}X^{a_2}_{q_2}|\Psi\rangle\right) \le d_2(q_2 \mid q_1), \tag{10g}$$

where Eqs. (10a), (10c), (10e) and (10g) follow from monotonicity of the trace distance and Eq. (10d) follows since $\boldsymbol{N}_{q_1q_2}$ is a PVM. Eqs. (10a)–(10g) imply

$$d_4(q_1,q_2) \le 2\big(d_2(q_1 \mid q_2) + d_2(q_2 \mid q_1)\big).$$

By Eq. (7),

$$\begin{aligned} d_4^2(q_1,q_2) &\le 4\big(d_2(q_1 \mid q_2) + d_2(q_2 \mid q_1)\big)^2 \\ &\le 8\big(d_2^2(q_1 \mid q_2) + d_2^2(q_2 \mid q_1)\big), \end{aligned}$$

where the last inequality follows from the Cauchy–Schwartz inequality. By using this, Eq. (9) is proved as follows:

$$\begin{aligned} \mathbf{E}_{q_1,q_2}[d_4^2(q_1,q_2)] &\le 8\,\mathbf{E}_{q_1,q_2}[d_2^2(q_1 \mid q_2) + d_2^2(q_2 \mid q_1)] \\ &\le 32\varepsilon_{\mathrm{cons}}. \end{aligned}$$

□

**Claim 4.** *Let $m \ge 1$, $t_1,\dots,t_m \in [Q]$ and $1 \le i \le m$. Then*

$$\frac{1}{2}\sum_{z_1,\dots,z_m\in A}\left|\|X^{z_m}_{t_m}X^{z_{m-1}}_{t_{m-1}}\cdots X^{z_1}_{t_1}|\Psi\rangle\|^2 - \|X^{z_m}_{t_m}\cdots X^{z_{i+1}}_{t_{i+1}}X^{z_{i-1}}_{t_{i-1}}\cdots X^{z_1}_{t_1}X^{z_i}_{t_i}|\Psi\rangle\|^2\right| \le 2\sum_{1\le j\le i-1}d_1(t_j) + \sum_{1\le j\le i-1}d_4(t_i,t_j).$$

*Proof.* We prove the claim by hybrid argument. For $1 \le j \le m$ and $z \in A^m$, let $X_j^z = X_{t_j}^{z_j}$ and $Y_j^z = Y_{t_j}^{z_j}$. For $1 \le j \le j' \le m$, let $X_{[j',j]}^z = X_{j'}^z X_{j'-1}^z \cdots X_j^z$ and $Y_{[j,j']}^z = Y_j^z Y_{j+1}^z \cdots Y_{j'}^z$. For $j' = j-1$, we let $X_{[j',j]}^z = Y_{[j,j']}^z = I$. For $1 \le j \le i$, we define probability distributions $p_j$ and $p'_j$ on $A^m$ by

$$p_j(z) = \|X_{[m,j]}^z Y_{[1,j-1]}^z |\Psi\rangle\|^2,$$
$$p'_j(z) = \|X_{[m,i+1]}^z X_{[i-1,j]}^z Y_{[1,j-1]}^z X_i^z |\Psi\rangle\|^2.$$

For $j = 1$ and $j = i$, these probability distributions are written as

$$p_1(z) = \|X_{[m,1]}^z |\Psi\rangle\|^2,$$
$$p'_1(z) = \|X_{[m,i+1]}^z X_{[i-1,1]}^z X_i^z |\Psi\rangle\|^2,$$

and

$$p_i(z) = \|X_{[m,i]}^z Y_{[1,i-1]}^z |\Psi\rangle\|^2,$$
$$p'_i(z) = \|X_{[m,i+1]}^z Y_{[1,i-1]}^z X_i^z |\Psi\rangle\|^2.$$

Note that $p_i(z) = p'_i(z)$ since $X_i^z$ commute with $Y_{j'}^z$ for $1 \le j' \le i-1$.

By using $p_j$ and $p'_j$, the inequality to prove can be written as

$$D(p_1, p'_1) \le 2 \sum_{1 \le j \le i-1} d_1(t_j) + \sum_{1 \le j \le i-1} d_4(t_j, t_i).$$

We prove this by proving

$$D(p_j, p_{j+1}) \le d_1(t_j), \qquad 1 \le j \le i-1, \tag{11}$$
$$D(p'_{j+1}, p'_j) \le d_1(t_j) + d_4(t_i, t_j), \quad 1 \le j \le i-1. \tag{12}$$

We prove Eq. (11). By monotonicity of the trace distance,

$$\frac{1}{2} \sum_{z \in A^m} \big| \|X_{[m,j+1]}^z Y_{[1,j-1]}^z X_j^z |\Psi\rangle\|^2 - \|X_{[m,j+1]}^z Y_{[1,j-1]}^z Y_j^z |\Psi\rangle\|^2 \big| \le d_1(t_j). \tag{13}$$

The second term inside the absolute value is equal to $p_{j+1}(z)$. The first term is equal to $p_j(z)$ since $X_j^z$ commute with $Y_{j'}^z$ for $1 \le j' \le j-1$. Therefore, Eq. (13) is identical to Eq. (11).

To prove Eq. (12), let $Z = X_{[m,i+1]}^z X_{[i-1,j+1]}^z Y_{[1,j-1]}^z$. Then by monotonicity of the trace distance,

$$\frac{1}{2} \sum_{z \in A^m} \big| \|Z X_i^z Y_j^z |\Psi\rangle\|^2 - \|Z X_i^z X_j^z |\Psi\rangle\|^2 \big| \le d_1(t_j), \tag{14}$$
$$\frac{1}{2} \sum_{z \in A^m} \big| \|Z X_i^z X_j^z |\Psi\rangle\|^2 - \|Z X_j^z X_i^z |\Psi\rangle\|^2 \big| \le d_4(t_i, t_j). \tag{15}$$

Note that $X_i^z$ commute with $Y_j^z$. This implies that $p'_{j+1}(z) = \|Z X_i^z Y_j^z |\Psi\rangle\|^2$. Since $X_j^z$ commute with $Y_{j'}^z$ for $j+1 \le j' \le m$, $\|Z X_j^z X_i^z |\Psi\rangle\|^2 = p'_j(z)$. Eq. (12) is obtained by adding Eqs. (14) and (15).

By summing up Eqs. (11) and (12) over $1 \le j \le i-1$,

$$D(p_1, p'_1) \le 2 \sum_{1 \le j \le i-1} d_1(t_j) + \sum_{1 \le j \le i-1} d_4(t_i, t_j). \qquad \square$$

Now let $q_1, q_2, q_3 \in [Q]$. We start with $\theta(\Pi) = \|X_Q^{\Pi[Q]} X_{Q-1}^{\Pi[Q-1]} \cdots X_1^{\Pi[1]} |\Psi\rangle\|^2$ and bring $X_{q_1}^{\Pi[q_1]}$, $X_{q_2}^{\Pi[q_2]}$ and $X_{q_3}^{\Pi[q_3]}$ to the nearest to $|\Psi\rangle$ by applying Claim 4 three times. Recall that questions $1, \dots, Q$ are sorted in the descending order of their marginal probabilities. If we write the $Q-3$ elements of $[Q] \setminus \{q_1, q_2, q_3\}$ as $v_1, \dots, v_{Q-3}$ in the same order as $1, \dots, Q$, we obtain

$$\frac{1}{2} \sum_{\Pi \in A^Q} \left| \theta(\Pi) - \|X_{v_{Q-3}}^{\Pi[v_{Q-3}]} X_{v_{Q-2}}^{\Pi[v_{Q-2}]} \cdots X_{v_1}^{\Pi[v_1]} X_{q_1}^{\Pi[q_1]} X_{q_2}^{\Pi[q_2]} X_{q_3}^{\Pi[q_3]} |\Psi\rangle\|^2 \right| \leq \sum_{1 \leq i \leq 3} \left( 2 \sum_{1 \leq q' \leq q_i - 1} d_1(q') + \sum_{1 \leq q' \leq q_i - 1} d_4(q_i, q') \right).$$

Note that

$$\sum_{\substack{\Pi \in A^Q \\ \Pi[q_i] = a_i \ (i=1,2,3)}} \|X_{v_{Q-3}}^{\Pi[v_{Q-3}]} X_{v_{Q-2}}^{\Pi[v_{Q-2}]} \cdots X_{v_1}^{\Pi[v_1]} X_{q_1}^{\Pi[q_1]} X_{q_2}^{\Pi[q_2]} X_{q_3}^{\Pi[q_3]} |\Psi\rangle\|^2 = \|X_{q_1}^{a_1} X_{q_2}^{a_2} X_{q_3}^{a_3} |\Psi\rangle\|^2,$$

and therefore,

$$\frac{1}{2} \sum_{a_1, a_2, a_3 \in A} \left| \theta(a_1, a_2, a_3 \mid q_1, q_2, q_3) - \|X_{q_1}^{a_1} X_{q_2}^{a_2} X_{q_3}^{a_3} |\Psi\rangle\|^2 \right| \leq \sum_{1 \leq i \leq 3} \left( 2 \sum_{1 \leq q' \leq q_i - 1} d_1(q') + \sum_{1 \leq q' \leq q_i - 1} d_4(q_i, q') \right). \tag{16}$$

By monotonicity of the trace distance,

$$\begin{aligned}
\frac{1}{2} \sum_{a_1, a_2, a_3 \in A} \left| \|X_{q_1}^{a_1} X_{q_2}^{a_2} X_{q_3}^{a_3} |\Psi\rangle\|^2 - \|X_{q_1}^{a_1} X_{q_2}^{a_2} Y_{q_3}^{a_3} |\Psi\rangle\|^2 \right| &\leq d_1(q_3), \\
\frac{1}{2} \sum_{a_1, a_2, a_3 \in A} \left| \|Y_{q_3}^{a_3} X_{q_1}^{a_1} X_{q_2}^{a_2} |\Psi\rangle\|^2 - \|Y_{q_3}^{a_3} X_{q_1}^{a_1} Y_{q_2}^{a_2} |\Psi\rangle\|^2 \right| &\leq d_1(q_2), \\
\frac{1}{2} \sum_{a_1, a_2, a_3 \in A} \left| \|Y_{q_3}^{a_3} Y_{q_2}^{a_2} X_{q_1}^{a_1} |\Psi\rangle\|^2 - \|Y_{q_3}^{a_3} Y_{q_2}^{a_2} Y_{q_1}^{a_1} |\Psi\rangle\|^2 \right| &\leq d_1(q_1),
\end{aligned}$$

which implies

$$\frac{1}{2} \sum_{a_1, a_2, a_3 \in A} \left| \|X_{q_1}^{a_1} X_{q_2}^{a_2} X_{q_3}^{a_3} |\Psi\rangle\|^2 - \|Y_{q_3}^{a_3} Y_{q_2}^{a_2} Y_{q_1}^{a_1} |\Psi\rangle\|^2 \right| \leq d_1(q_1) + d_1(q_2) + d_1(q_3). \tag{17}$$

Similarly,

$$\begin{aligned}
&\frac{1}{2} \sum_{a_1, a_2, a_3 \in A} \left| \|Y_{q_3}^{a_3} Y_{q_2}^{a_2} Y_{q_1}^{a_1} |\Psi\rangle\|^2 - \|M_{q_1 q_2 q_3}^{a_1 a_2 a_3} |\Psi\rangle\|^2 \right| \\
&= \frac{1}{2} \sum_{a_1, a_2, a_3 \in A} \left| \|Y_{q_3}^{a_3} Y_{q_2}^{a_2} Y_{q_1}^{a_1} |\Psi\rangle\|^2 - \|M_{q_1|q_1 q_2 q_3}^{a_1} M_{q_2|q_1 q_2 q_3}^{a_2} M_{q_3|q_1 q_2 q_3}^{a_3} |\Psi\rangle\|^2 \right| \\
&\leq \sum_{1 \leq i \leq 3} d_3(q_i \mid q_1 q_2 q_3),
\end{aligned} \tag{18}$$

where the first equality is since $\boldsymbol{M}_{q_1 q_2 q_3}$ is a PVM.

By Eqs. (16), (17) and (18),

$$\frac{1}{2}\sum_{a_1,a_2,a_3\in A}\left|\theta(a_1,a_2,a_3 \mid q_1,q_2,q_3) - \|M_{q_1q_2q_3}^{a_1a_2a_3}|\Psi\rangle\|^2\right| \le d(q_1,q_2,q_3),$$

where

$$\begin{aligned} d(q_1,q_2,q_3) &= \sum_{1\le i\le 3}\left(2\sum_{1\le q'\le q_i-1} d_1(q') + \sum_{1\le q'\le q_i-1} d_4(q_i,q') + d_1(q_i) + d_3(q_i \mid q_1q_2q_3)\right) \\ &\le \sum_{1\le i\le 3}\left(2\sum_{1\le q'\le q_i} d_1(q') + \sum_{1\le q'\le q_i-1} d_4(q_i,q') + d_3(q_i \mid q_1q_2q_3)\right). \end{aligned}$$

By the linearity of expectation,

$$\mathbf{E}_{(q_1,q_2,q_3)}[d(q_1,q_2,q_3)] \le 6\,\mathbf{E}_q\left[\sum_{1\le q'\le q} d_1(q')\right] + 3\,\mathbf{E}_q\left[\sum_{1\le q'\le q-1} d_4(q,q')\right] + 3\,\mathbf{E}_{(q_1,q_2,q_3)}\,\mathbf{E}_i[d_3(q_i \mid q_1q_2q_3)]. \tag{19}$$

We bound each term of the right-hand side of Eq. (19). As for the first term,

$$\begin{aligned} \left(\mathbf{E}_q\left[\sum_{1\le q'\le q} d_1(q')\right]\right)^2 &= \left(\sum_{1\le q'\le q\le Q}\pi(q)d_1(q')\right)^2 \\ &\le \left(\sum_{1\le q'\le q\le Q}\pi(q')d_1(q')\right)^2 && (q'\le q \implies \pi(q')\ge\pi(q)) \\ &\le Q^2\left(\sum_{1\le q'\le Q}\pi(q')d_1(q')\right)^2 = Q^2\big(\mathbf{E}_{q'}[d_1(q')]\big)^2 \\ &\le Q^2\,\mathbf{E}_{q'}[d_1^2(q')] \\ &\le 2Q^2\varepsilon_{\mathrm{cons}}. && \text{(by Claim 3)} \end{aligned}$$

As for the second term,

$$\begin{aligned} \left(\mathbf{E}_q\left[\sum_{1\le q'\le q-1} d_4(q,q')\right]\right)^2 &= \left(\sum_{1\le q'<q\le Q}\pi(q)d_4(q,q')\right)^2 \\ &\le \left(\sum_{1\le q'<q\le Q}\sqrt{\pi(q)\pi(q')}\,d_4(q,q')\right)^2 \\ &\qquad (q'\le q \implies \pi(q')\ge\pi(q) \implies \sqrt{\pi(q)\pi(q')}\ge\pi(q)) \\ &\le \frac{1}{4}\left(\sum_{1\le q,q'\le Q}\sqrt{\pi(q)\pi(q')}\,d_4(q,q')\right)^2 \\ &\le \frac{Q^2}{4}\sum_{1\le q,q'\le Q}\pi(q)\pi(q')d_4^2(q,q') && \text{(by the Cauchy–Schwartz inequality)} \end{aligned}$$

$$
\begin{aligned}
&= \frac{Q^2}{4}\,\mathbf{E}_{q,q'}[d_4^2(q,q')] \\
&\le 8Q^2\varepsilon_{\mathrm{cons}}. && \text{(by Claim 3)}
\end{aligned}
$$

As for the third term,

$$
\begin{aligned}
\left(\mathbf{E}_{(q_1,q_2,q_3)}\,\mathbf{E}_i[d_3(q_i \mid q_1,q_2,q_3)]\right)^2 &\le \mathbf{E}_{(q_1,q_2,q_3)}\,\mathbf{E}_i[d_3^2(q_i \mid q_1,q_2,q_3)] \\
&\le 2\varepsilon_{\mathrm{cons}}. && \text{(by Claim 3)}
\end{aligned}
$$

Therefore,

$$
\begin{aligned}
\mathbf{E}_{(q_1,q_2,q_3)}[d(q_1,q_2,q_3)] &\le 6\sqrt{2}\,Q\sqrt{\varepsilon_{\mathrm{cons}}} + 3\sqrt{8}\,Q\sqrt{\varepsilon_{\mathrm{cons}}} + 3\sqrt{2}\sqrt{\varepsilon_{\mathrm{cons}}} \\
&= (12\sqrt{2}\,Q + 3\sqrt{2})\sqrt{\varepsilon_{\mathrm{cons}}} \\
&\le 15\sqrt{2}\,Q\sqrt{\varepsilon_{\mathrm{cons}}}.
\end{aligned}
$$

By Eq. (5),

$$
1 - w(G) - \varepsilon_{\mathrm{sim}} \le |1 - w(G) - \varepsilon_{\mathrm{sim}}| \le \mathbf{E}_{(q_1,q_2,q_3)}[d(q_1,q_2,q_3)] \le 15\sqrt{2}\,Q\sqrt{\varepsilon_{\mathrm{cons}}},
$$

which implies

$$
\varepsilon_{\mathrm{sim}} \ge 1 - w(G) - 15\sqrt{2}\,Q\sqrt{\varepsilon_{\mathrm{cons}}}.
$$

Since $\varepsilon \ge \varepsilon_{\mathrm{sim}}$ and $\varepsilon \ge \varepsilon_{\mathrm{cons}}$,

$$
\sqrt{\varepsilon} \ge \varepsilon \ge \varepsilon_{\mathrm{sim}} \ge 1 - w(G) - 15\sqrt{2}\,Q\sqrt{\varepsilon_{\mathrm{cons}}} \ge 1 - w(G) - 15\sqrt{2}\,Q\sqrt{\varepsilon},
$$

which implies

$$
\varepsilon \ge \left(\frac{1 - w(G)}{1 + 15\sqrt{2}\,Q}\right)^2 \ge \frac{(1 - w(G))^2}{(1 + 15\sqrt{2})^2 Q^2}.
$$

This means that Lemma 8 holds with $c = 1/(1 + 15\sqrt{2})^2$.

*Remark.* Kempe, Kobayashi, Matsumoto, Toner, and Vidick [20] essentially uses nonadaptive two-query PCP systems with a nonboolean alphabet as a base case. We use nonadaptive three-query PCP systems with the boolean alphabet to make the answers in the resulting games shorter. Note that this difference is not essential unless we are not concerned with the answer length: if we allow four-bit (or more precisely, two-trit) answers in the resulting games, then we could use nonadaptive two-query PCP systems with the ternary alphabet as well.

Actually, if we define "nonadaptive $t$-query PCP games" for general $t \in \mathbb{N}$ analogously to the definition of nonadaptive three-query PCP games, it is easy to generalize oracularization with a dummy question to nonadaptive $t$-query PCP games, and Lemma 8 also applies by changing the coefficients $1/3$ and $c$ in the statement depending on $t$.